\documentclass[aps,prx,superscriptaddress,twocolumn]{revtex4-1}

\usepackage{graphicx}
\usepackage{amsmath}
\usepackage{amssymb}
\usepackage{color}
\definecolor{darkblue}{rgb}{0.,0.,0.5}
\definecolor{darkred}{rgb}{0.7,0.,0.}
\usepackage[colorlinks=true,linkcolor=blue,citecolor=blue,urlcolor=blue]{hyperref}
\usepackage{soul}
\usepackage{bbm}
\usepackage{commath}
\usepackage{mathptmx}
\usepackage{enumerate}
\usepackage{mathtools}
\usepackage{times}

\DeclareMathAlphabet{\pazocal}{OMS}{zplm}{m}{n}

\DeclareMathOperator{\tr}{tr}
\DeclareMathOperator{\diag}{diag}

\newcommand{\comments}[1]{}
\newcommand{\bea}{\begin{eqnarray}}
\newcommand{\eea}{\end{eqnarray}}

\makeatletter
\DeclareFontFamily{OMX}{MnSymbolE}{}
\DeclareSymbolFont{MnLargeSymbols}{OMX}{MnSymbolE}{m}{n}
\SetSymbolFont{MnLargeSymbols}{bold}{OMX}{MnSymbolE}{b}{n}
\DeclareFontShape{OMX}{MnSymbolE}{m}{n}{
	<-6>  MnSymbolE5
	<6-7>  MnSymbolE6
	<7-8>  MnSymbolE7
	<8-9>  MnSymbolE8
	<9-10> MnSymbolE9
	<10-12> MnSymbolE10
	<12->   MnSymbolE12
}{}
\DeclareFontShape{OMX}{MnSymbolE}{b}{n}{
	<-6>  MnSymbolE-Bold5
	<6-7>  MnSymbolE-Bold6
	<7-8>  MnSymbolE-Bold7
	<8-9>  MnSymbolE-Bold8
	<9-10> MnSymbolE-Bold9
	<10-12> MnSymbolE-Bold10
	<12->   MnSymbolE-Bold12
}{}

\newcommand{\ignore}[1]{}
\newcommand{\nobibentry}[1]{{\let\nocite\ignore\bibentry{#1}}}

\newcommand{\bibfnamefont}[1]{#1}
\newcommand{\bibnamefont}[1]{#1}

\newcommand{\pO}{\pazocal{O}}
\newcommand{\tJ}{\tilde{J}}
\newcommand{\tom}{\tilde{\omega}}
\newcommand{\tT}{\tilde{T}}

\begin{document}

\title{Measuring the temperature of cold many-body quantum systems}

\date{\today}

\author{Karen V. Hovhannisyan}
\affiliation{Department of Physics and Astronomy, Ny Munkegade 120, 8000 Aarhus, Denmark}
\email{karen@phys.au.dk}

\author{Luis A. Correa}
\affiliation{School of Mathematical Sciences and Centre for the Mathematics and Theoretical Physics of Quantum Non-Equilibrium Systems, The University of Nottingham, University Park, Nottingham NG7 2RD, United Kingdom}
\email{luis.correa@nottingham.ac.uk}

\begin{abstract}

Precise low-temperature thermometry is a key requirement for virtually any quantum technological application. Unfortunately, as the temperature $ T $ decreases, the errors in its estimation diverge very quickly. In this paper, we determine exactly how quickly this may be. We rigorously prove that the ``conventional wisdom'' of low-$T$ thermometry being exponentially inefficient, is limited to local thermometry on translationally invariant systems with short-range interactions, featuring a non-zero gap above the ground state. This result applies very generally to spin and harmonic lattices. On the other hand, we show that a power-law-like scaling is the hallmark of local thermometry on \textit{gapless} systems. Focusing on thermometry on one node of a harmonic lattice, we obtain valuable physical insight into the switching between the two types of scaling. In particular, we map the problem to an equivalent setup, consisting of a Brownian thermometer coupled to an equilibrium reservoir. This mapping allows us to prove that, surprisingly, the relative error of local thermometry on gapless harmonic lattices does not diverge as $T\rightarrow 0$; rather, it saturates to a constant. As a useful by-product, we prove that the low-$T$ sensitivity of a harmonic probe arbitrarily strongly coupled to a bosonic reservoir by means of a generic Ohmic interaction, always scales as $T^2$ for $T\rightarrow 0$. Our results thus identify the energy gap between the ground and first excited states of the \textit{global} system as the key parameter in \textit{local} thermometry, and ultimately provide clues to devising practical thermometric strategies deep in the ultra-cold regime.

\end{abstract}

\maketitle

\section{Introduction} \label{sec:introduction}

Sub-micron thermometry has developed into an experimentally mature discipline, thus enabling high-resolution temperature measurements with nanometer-sized probes \cite{carlos2015thermometry}, and even individual \textit{quantum thermometers} \cite{neumann2013high,kucsko2013nanometre}. Although measuring low temperatures with high precision is notoriously hard \cite{Wu_1998, Correa_2015, mehboudi2015ultracold, De_Pasquale_2016, Paris_2016}, the quest to produce accurate quantum thermometers fit for use in the ultra-cold regime is strongly driven by their potential applications in, e.g., quantum information processing \cite{halbertal2016nanoscale}. Understanding the origin of the severe fundamental limitations that hinder precise low-temperature thermometry is thus essential for future technological developments.

The temperature $ T $ of an equilibrium quantum system can be accurately calculated from a large collection of energy measurements. Indeed, knowing its spectrum and assuming that the populations in the energy basis follow a Boltzmann distribution allows to build a maximum likelihood estimator for $ T $ \cite{kay1993fundamentals}. As it turns out, such strategy is optimal, in the sense that it achieves the smallest possible mean squared error on the estimated temperature \cite{Mandelbrot_1956, Correa_2015, Marzolino_2013}. Indeed, energy measurements allow to \textit{saturate} the quantum Cram\'er-Rao inequality $ (\delta T)^2 \geq 1/(M \pazocal{F}_T) $ \cite{PhysRevLett.72.3439} for temperature estimation on any equilibrium system. In other words, the inverse of the mean squared error of the final estimate (normalized by the length $ M $ of dataset of measurement outcomes) converges to the so-called quantum Fisher information (QFI) $ \pazocal{F}_T \overset{M\rightarrow\infty}{=} (\delta T)^{-2}/M $, which is, in our case, a quantifier of ``thermal sensitivity''. 

In turn, the QFI relates to the heat capacity $ C(T) \coloneqq d\langle H \rangle/dT $ of the equilibrium system, as $ \pazocal{F}_T = C(T)/T^2 $ \cite{Correa_2015, Reeb_2015}. Here, $ H $ is the system Hamiltonian, $ \langle \cdots \rangle $ denotes thermal averaging, and we have adopted units in which $ \hbar = k_B = 1 $. Very generally, the heat capacity of a  finite-size quantum system at equilibrium decays exponentially fast to zero \cite{einstein1906plancksche} in the limit $ T\rightarrow 0 $, i.e., $ C(T) \sim \pO(e^{-\Delta/T}) $, where $ \Delta $ stands for the energy gap between ground and first excited states. Hence, thermometry on a finite equilibrium quantum system becomes exponentially inefficient at low temperatures \cite{Correa_2015, Correa_2017}.

As the system scales up in size, its heat capacity and hence, also its thermal sensitivity, grows \textit{extensively} (see Appendix \ref{app:heat_capacity}). However, benefiting from size-scaling would require making generally unfeasible global multi-particle measurements, while strongly perturbing the system. Although thermometrically useful non-demolition global measurements can sometimes be implemented \cite{eckert2007quantum, Mehboudi_2016, sabin2014impurities, mehboudi2015ultracold}, these schemes rely on measurements of additive quantities and thus cannot be optimal for strongly interacting systems. This motivates the development of minimally invasive ``local'' thermometric strategies, aimed at inferring the global temperature from measurements on an accessible small fraction of the system \cite{De_Pasquale_2016, De_Palma_2017}.

The marginal state of the accessible part may deviate significantly from thermal equilibrium when it interacts strongly with the rest of the system. This is due to the large correlations built up between the two, especially at low temperatures. As a result, the internal interaction strength may be used to achieve some quantitative improvement over the local equilibrium situation. Nonetheless, whenever the global system is gapped, translationally invariant, short-range interacting, and non-critical, the exponential scaling at low temperatures is inescapable, as we will prove in Sec.~\ref{sec:exponential_inefficiency}. 

Alternatively, one may adopt an open-system perspective to gain additional insight into the problem of local thermometry \cite{Correa_2017}, since the dissipative dynamics of an individual quantum \textit{probe} coupled to an initially thermal and large \textit{sample}, converges to a global equilibrium state \cite{Subasi_2012}. In this sense, ``probe'' and ``sample'' match the ``accessible'' and ``inaccessible'' parts of the many-body system in the setting outlined above. In particular, linear open quantum systems are especially interesting, as the reduced steady state of the probe can be found \textit{exactly} \cite{Weiss_1999, Subasi_2012}. 

Following this approach, we have been able to establish in Ref.~\cite{Correa_2017} that $ \pazocal{F}_T\sim T^{2} $ in the limit $ T \rightarrow 0 $, for the simplest model of Brownian motion, i.e., a single harmonic thermometer coupled to an initially thermal bosonic reservoir \cite{caldeira1983path, Riseborough_1985, grabert1984quantum}. Our results followed from the exact analytical steady-state solution, and indicated that such power-law-like scaling holds for various instances of Ohmic and super-Ohmic spectral densities. In Sec.~\ref{sec:power-law-like} and Appendix \ref{app:Ohmic_CL}, we will rigorously prove that a power-law-like scaling is indeed a signature of the ubiquitous Ohmic dissipation scheme: On probes with finite \textit{bare} trapping frequency $ \omega_0 > 0 $ we find a quadratic asymptotic scaling $ \pazocal{F}_T\sim T^2 $, while the diverging behavior $ \pazocal{F}_T \sim 1/T^2 $ is indicative of thermometers for which $ \omega_0\rightarrow 0 $.

This exact result seems to be in stark contradiction with the discussion above: How can a (tiny) subsystem display a power-law-like thermal sensitivity if global thermometry on the whole should be exponentially inefficient? The answer is that this model is \textit{gapless}, and hence fundamentally different from gapped systems. Namely, gapless systems are necessarily infinite, i.e., large enough to justify taking the thermodynamic limit. In addition, they do not have a parameter akin to the spectral gap $ \Delta $, that would set an energy scale. 

In order to gain physical insight, we turn to quantum harmonic systems. Specifically, we will consider 1D translationally invariant harmonic chains (TIHCs) with linear (but not necessarily short-range) interactions, prepared in thermal equilibrium. It is easy to verify that local thermometry, on a single node of a gapped instance of the chain, is exponentially inefficient, while power-law-like behavior shows up when a vanishing gap is enforced. We then try to look at such TIHCs from an open-system perspective: We diagonalize the inaccessible part so as to bring the problem into an equivalent open-system-like ``star'' configuration, where the probe is surrounded by non-interacting peripheral sample modes. 

Crucially, we find that gapped chains map into open systems with unusual spectral densities, where the low frequency sample modes appear decoupled from the probe. On the other hand, gapless chains give rise to standard Ohmic spectral densities. Hence, a gapless TIHC maps into the paradigmatic Caldeira-Leggett model (CLM) \cite{caldeira1983path} in the thermodynamic limit. Conversely, we show that thermometry on the CLM is equivalent to local temperature measurements on a gapless TIHC. Quite intuitively, this open-system viewpoint indicates that the ability of the probe to detect near-ground-state temperature fluctuations critically depends on whether or not it is effectively coupled to the lowermost normal modes of the sample, which, in turn, are the only ones that may be thermally populated at very low $ T $ \cite{De_Pasquale_2016, Correa_2017}. 

Our findings imply that engineering the probe-sample coupling to guarantee a good thermal contact with the low-frequency modes is the key to precise low-temperature quantum thermometry. In this sense, reservoir engineering and dynamical control \cite{mukherjee2017high} could come to play a major role in practical technological applications.

\section{Results}\label{sec:results}

\subsection{Exponential inefficiency of local quantum thermometry in gapped lattice systems}\label{sec:exponential_inefficiency}

\begin{figure}
\centering
\includegraphics[width=0.4\textwidth]{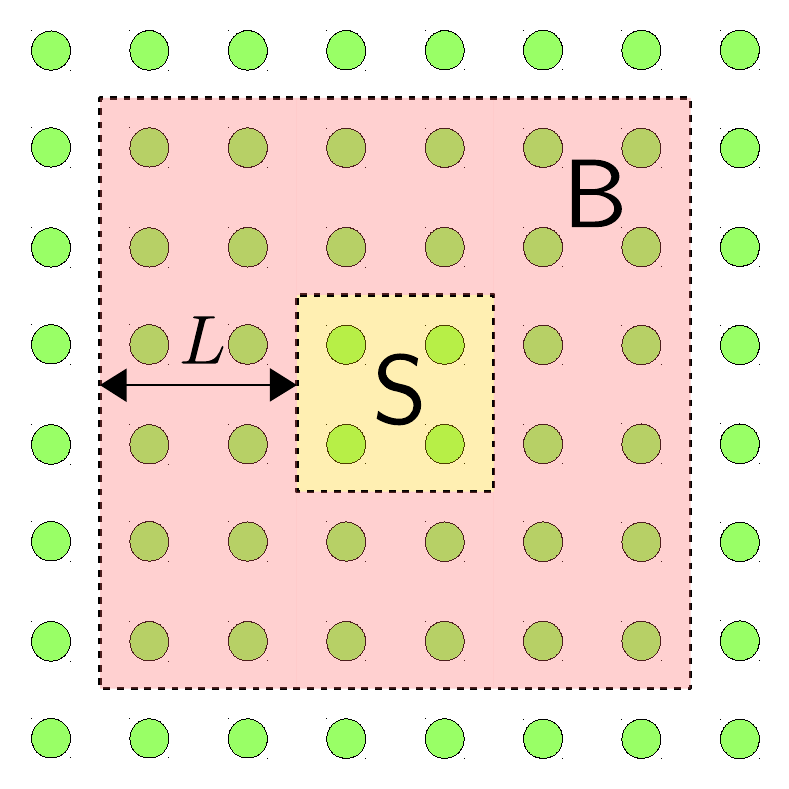}
\caption{(color online) Schematic diagram of a 2D translationally invariant lattice. The system $ S $ and the boundary $ B $ (of length $ L $) appear depicted in orange and red, respectively. Although not shown, the nodes (gray dots) of the lattice are are connected by short-ranged interactions. It is also assumed that the many-body system is away from criticality.}
\label{fig:lattice}
\end{figure}

Here, we shall consider the problem of local thermometry on an equilibrium lattice of identical harmonic oscillators or spins at temperature $ T $. To be precise, we will look at arbitrarily large translationally-invariant gapped lattices featuring arbitrarily strong but finite-range interactions. We are interested in the low-temperature regime, therefore, even if the parameters of the Hamiltonian are such that the lattice undergoes a (classical) second-order phase transition at some non-zero temperature $T_\text{crit}$, we consider only $T\neq T_\text{crit}$. This does not limit the generality of our low-temperature results as they clearly apply for all $T < T_\text{crit}$.

Our task will be to infer the global temperature $ T $ from local observations on the (non-equilibrium) marginal state of a finite-size subsystem $ S $. Two conflicting factors are at play here: On the minus side, the temperatures are low and $ S $ is of finite size, which makes us expect exponentially inefficient thermometry. However, $ S $ is \textit{strongly correlated} with the rest of the lattice, which due to its size, has an overall large heat capacity. We will show that the first factor nevertheless prevails and the local thermal sensitivity decays exponentially, namely as
\bea \label{no15}
\pazocal{F}_T \leq \pazocal{O}(1)
e^{-\beta\Delta} \quad \text{(for}\;\; \beta\Delta \gg 1\text{)}, 
\eea
where the QFI $ \pazocal{F}_T $ is adopted as the figure of merit and $\beta \coloneqq 1/T $. The QFI can be formally defined as
\begin{equation}
\pazocal{F}_T = -2\lim_{\delta\rightarrow 0}\frac{\partial^2 \mathbb{F}(\rho_T,\rho_{T+\delta})}{\partial\delta^2},
\label{eq:qfi}    
\end{equation}
where $ \rho_\theta $ denotes the marginal of $ S $ when the global lattice is at temperature $ \theta $ and $ \mathbb{F}(\rho,\sigma) \coloneqq (\tr\sqrt{\sqrt{\rho}\sigma\sqrt{\rho}})^2 $ is the Uhlmann fidelity \cite{uhlmann1976transition} between states $ \rho $ and $ \sigma $. Note how, already from its definition, it is intuitively clear that $ \pazocal{F}_T $ gauges the \textit{responsiveness} of the probe to small fluctuations in the sample temperature. For convenience, we shall cast $ \pazocal{F}_T $ in the equivalent form (cf. Appendix \ref{app:from_qfi_to_qfialt})
\begin{equation}
\pazocal{F}_T = 4\lim_{\delta \to 0}\frac{1-\mathbb{F}(\rho_T,\rho_{T+\delta})}{\delta^2},    
\label{eq:qfi_alt}
\end{equation} 
so that $ \mathbb{F}(\rho_T,\rho_{T+\delta}) =  1 - \frac{1}{4}\pazocal{F}_T\delta^2 + \pazocal{O}(\delta^3) $ and hence, the Bures distance $ d_B^2(\rho_T,\rho_{T+\delta}) \coloneqq 2 \big(1-\sqrt{\mathbbm{F}(\rho_T,\rho_{T+\delta})}\big) $ becomes
\begin{equation}
d_B^2(\rho_T,\rho_{T+\delta}) = \frac14\pazocal{F}_T\,\delta^2 + \pazocal{O}(\delta^3).
\label{eq:Bures_Fisher}
\end{equation}
In what follows, we will use general arguments on locality of temperature \cite{Garcia-Saez_2009, Kliesch_2014, Hernandez-Santana_2015} to bound the left-hand side of Eq.~\eqref{eq:Bures_Fisher} from above and thus, extract the low--$ T $ scaling of $ \pazocal{F}_T $. 

Let us denote the number of sites, i.e., the ``size'' of the system, by $ L_S $ and introduce a \textit{boundary region} around it, of size $ L $ (see Fig.~\ref{fig:lattice}). Also let $ \tau_T(L_S+L) $ be a thermal state at temperature $ T $ defined from the local Hamiltonian of ``system + boundary'' and $ \rho_T(L) \coloneqq \tr_B\tau(L_S + L)$, the reduction of $ \tau_T(L_S + L) $ within $ S $. By definition, one thus has $ \rho_T(L)\overset{L\rightarrow\infty}{=}\rho_T $. From the triangle inequality, it follows that
\begin{align}
d_B(\rho_T,\rho_{T+\delta}) &\leq d_B(\rho_T(L),\rho_{T+\delta}(L)) \nonumber\\ 
&+ d_B(\rho_{T},\rho_{T}(L)) + d_B(\rho_{T+\delta},\rho_{T+\delta}(L)).
\label{eq:bures1}
\end{align}
We can now use the data-processing inequality for the Bures distance \cite{mikeike} to bound the first term of the right-hand side as $ d_B(\rho_T(L),\rho_{T+\delta}(L)) \leq d_B(\tau_T(L_S+L),\tau_{T+\delta}(L_S+L)) $. According to Eq.~\eqref{eq:Bures_Fisher}, the latter can be cast as $ \frac12\sqrt{\pazocal{F}_T(L_S + L)}\delta + \pazocal{O}(\delta^2) $ and, since $ \tau_T(L_S+L) $ is a thermal state, we may additionally exploit the relation $ \pazocal{F}_T(L_S + L) = C_{S + B}(T)/T^2 $, where $ C_{S + B}(T) $ is the heat capacity of the ``system + boundary'' composite. Hence, we finally arrive at
\begin{align}
d_B(\rho_T,\rho_{T+\delta}) &\leq \frac{\sqrt{C_{S+B}(T)}}{2T}\,\delta 
+ d_B(\rho_{T},\rho_{T}(L))  \nonumber \\
&+ d_B(\rho_{T+\delta},\rho_{T+\delta}(L)) + \pazocal{O}(\delta^2). 
\label{eq:bures2}
\end{align}

Focusing now on the locally finite case (i.e., each node has a finite-dimensional Hilbert state space) and, given that we work with finite-range-interacting systems away from criticality, we may approximate $ \rho_T = \rho_T(\infty) $ by $ \rho_T(L) $ for finite but large boundary \cite{Garcia-Saez_2009, Kliesch_2014, Hernandez-Santana_2015}. It has been shown in Ref.~\cite{Hernandez-Santana_2015} that, for such systems in 1D,
\begin{equation}
\mathbb{F}(\rho_T(L),\rho_T) = 1 - \pazocal{O}(1)e^{-L/\xi(T)},
\label{eq:fidelity_decay}
\end{equation}
where $\xi(T)$ is a monotonic function of the correlation length of the infinite chain, and, since the system is away from criticality, $ \xi(T) $ is also a regular function of $ T $, even when $ T \to 0 $. In terms of Bures distance, we can thus write
\begin{equation} 
d_B(\rho_T(L),\rho_T) = \pazocal{O}(1)e^{-\frac{L}{2\xi(T)}}.
\label{eq:bures4}
\end{equation}

Let us define $ \xi \coloneqq \max\{ \xi(T), \xi(T+\delta) \} $, so that Eq.~\eqref{eq:bures2} can be cast as
\begin{equation}
d_B(\rho_T,\rho_{T+\delta}) \leq \frac{\sqrt{C_{S+B}(T)}}{2T}\,\delta + \pazocal{O}(1)e^{-\frac{L}{2\xi}} + \pazocal{O}(\delta^2).
\label{eq:bures5}
\end{equation}

In Appendix \ref{app:heat_capacity}, we argue that the heat capacity of gapped locally finite translationally invariant lattices with nearest-neighbour interactions, and that of harmonic, translationally invariant, and not necessarily short-range interacting $N$-body systems, scales as 
\bea\label{eq:heat_text}
C_{S+B}(T)=\pO\left(N e^{-\beta\Delta}\right),
\eea
which leads us to
\bea \label{no10}
d_B(\rho_T,\rho_{T+\delta}) \leq \pazocal{O}(1)\left[ \sqrt{L_S+L}
\,e^{-\frac{\beta\Delta}{2}}\,\delta +  e^{-\frac{L}{2\xi}} + \delta^2 \right].~~~
\eea
Furthermore, in Appendix \ref{app:heat_capacity}, the scaling in Eq.~\eqref{eq:heat_text} is illustrated in the quantum Ising model.

Since we are interested in the low-temperature regime, we shall take the limit $ \beta\Delta\to\infty $. Therefore, we must ensure that $ \delta $ is much smaller than $ T $ (so that our Taylor expansions above make sense). Let us thus choose $\delta = \Delta\,e^{-\beta\Delta}$, which, as needed, satisfies $\delta/T=(\beta\Delta) e^{-\beta\Delta}\ll 1$. Furthermore, let $ L = 4\xi\beta\Delta $ [so that $ L \gg 1 $, as it was necessary for Eq.~\eqref{eq:bures4}]. Substituting these into \eqref{no10}, we arrive at
\bea \label{no14}
d_B(\rho_T,\rho_{T+\delta}) \leq \pazocal{O}(1) \! \left( 
e^{-\frac{3}{2}\beta\Delta} + \delta^2 \right) \! = \pazocal{O}(1)
e^{-\frac{3}{2}\beta\Delta},
\eea
which, due to Eq.~\eqref{eq:Bures_Fisher}, leaves us with Eq.~\eqref{no15}. We have thus shown that, very generally, local thermometry is exponentially inefficient at low temperatures in arbitrarily strongly but finite-range interacting lattices. This is one of our main results. We note that the proof can be readily extended to the cases when the interaction is not strictly of finite range but decays exponentially at large distances.

Although our proof is rigorous only in 1D, we expect Eq.~\eqref{no15} to be widely applicable also in higher dimensions. Indeed, except for Eq.~\eqref{eq:fidelity_decay}, all the steps in the proof hold in any spatial dimension. When it comes to Eq.~\eqref{eq:fidelity_decay}, it should generically apply to lattices with locally finite-dimensional Hilbert spaces, that are away from criticality, irrespective of spatial dimension \cite{Kliesch_2014}. Furthermore, results about approximating $\rho_T(\infty)$ with $\rho_T(L)$ \cite{Ferraro_2012}, about the relation between the spectral gap and exponential decay of correlations in generic harmonic lattices \cite{Cramer_2006}, and our Appendix \ref{app:heat_capacity}, strongly suggest that Eq.~\eqref{no15} should also be applicable in generic harmonic lattice systems in any dimension.

\subsection{Power-law-like sensitivity of a Brownian thermometer coupled to a sample through an Ohmic interaction}\label{sec:power-law-like}

\begin{figure}
\centering
\includegraphics[width=0.46\textwidth]{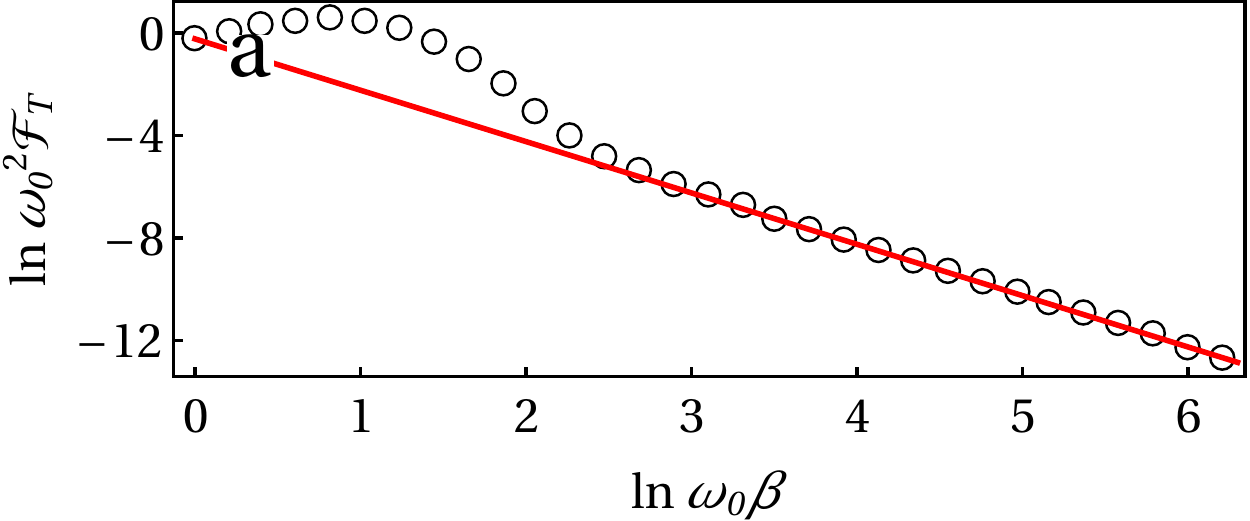}
\includegraphics[width=0.44\textwidth]{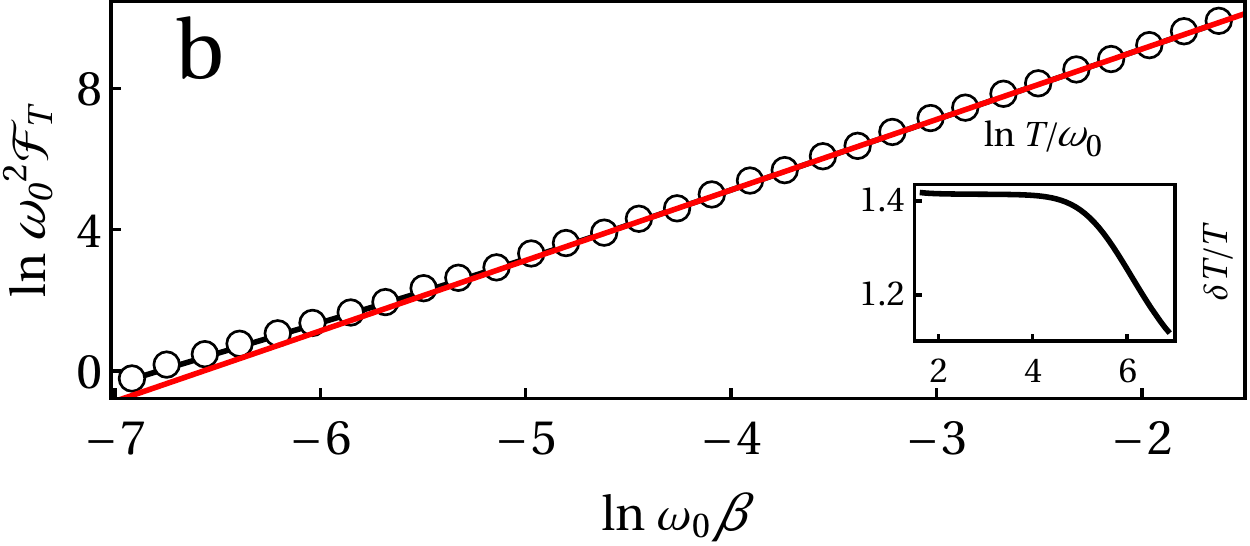}
\caption{(color online) \textbf{(a)} Log-log plog of the QFI as a function of the inverse temperature $ \beta = 1/T $ (open circles) for a probe with bare frequency $ \omega_0 > 0 $. The low-temperature asymptotic behavior $ \pazocal{F}_T \sim T^2 $ appears superimposed in red. \textbf{(b)} Same as in (a) but for a free probe ($ \omega_0 \rightarrow 0 $). In this case, the low--$ T $ scaling is $ \pazocal{F}_T\sim 1/T^2 $ (red). In (a) $ \omega_0 = 1 $, while in (b), $ \omega_0 = 10^{-3} $. \textbf{(inset)} Best-case relative error $ \delta T/T = 1/T\sqrt{M \pazocal{F}_T} $ for $ M = 1 $ as a function of $ T $ for the same parameters as in (b). In the two panels, we chose $ \gamma = 0.1 $ and $ \omega_c = 100 $. Note that, in both cases, the power-law-like scalings are maintained well beyond the bounds in Eqs.~\eqref{eq:scalings}.  
}
\label{fig:scaling_QLE}
\end{figure}

In this section, we will consider the low-temperature scaling of the sensitivity of a Brownian quantum particle. To that end, we shall adopt the quintessential Caldeira-Leggett model \cite{caldeira1983path, Weiss_1999}, consisting of a quantum harmonic thermometer of bare frequency $ \omega_0 $, linearly coupled to a bosonic reservoir (playing the role of the ``sample'') through an Ohmic interaction scheme. The sample will be initially prepared in thermal equilibrium at some unknown temperature $ T $ (to be measured). The probe, on the other hand, may be initialized in an arbitrary state, so long as it starts uncorrelated from the sample. The dissipative dynamics following from the thermal contact between probe and sample will bring the composite to a global equilibrium state \cite{Subasi_2012} at temperature $ T $. 

The Caldeira-Leggett Hamiltonian $ H_\text{CL} $, reads
\begin{align} \label{eq:Caldeira_Leggett}
H_\text{CL} = \frac{p_0^2}{2} + \frac12 (\omega_0^2+\omega_R^2)\frac{q_0^2}{2} &+  q_0 \sum_{\mu}  g_\mu q_\mu \nonumber\\  
&+  \sum_{\mu}  \left(\frac{p_\mu^2}{2} + \frac{q_\mu^2 \omega_\mu^2}{2} \right),
\end{align}
where $ (q_0,p_0) $ are the position and momentum quadratures of the probe, $ (q_\mu,p_\mu) $ are those of mode $ \omega_\mu $ in the sample, and the $ g_\mu $ stand for the probe-sample coupling strengths. The term $ \omega_R^2 \coloneqq \sum_{\mu} g_\mu^2/\omega_\mu^2 $ needs to be added ``by hand'' in order to ensure that $ H_\text{CL} $ is positive-definite. One neat way to understand the role of $ \omega_R $ is to write the effective potential ``felt'' by the Brownian particle \cite{Weiss_1999}. Notice that the choice $ q_\mu = q_{\mu,\rm min} \coloneqq - q_0 g_\mu/\omega_\mu^2 $ minimizes the potential energy contribution to Eq.~\eqref{eq:Caldeira_Leggett}, 
$V(q_0,\{ q_\mu \})$; that is, $ \partial_{q_{\mu}} V(q_0,\{ q_\mu \})\vert_{q_\mu = q_{\mu,\min}} = 0 $. Hence, the effective potential for the particle, $V_\text{eff}(q_0) = V(q_0,\{ q_{\mu,\min} \})$, writes as
\begin{eqnarray}
V_\text{eff}(q_0) = \frac{1}{2} q_0^2 \left(\omega_0^2 + \omega_R^2 - \sum_{\mu}\frac{g_\mu^2}{\omega_\mu^2} \right).
\end{eqnarray}
We thus see that the frequency renormalization exactly cancels the distortion on the potential of the particle due to its interaction with the sample. In particular, no matter how strong the probe-sample coupling might be, $ V_{\rm eff}(q_0) $ would never become \textit{inverted}. As a result, in order to model the situation $ \omega_0 = 0 $, what needs to be coupled to the sample is a Brownian particle \textit{trapped} at frequency $ \omega_R > 0 $ [cf. Appendix \ref{app:scaling_free}].

Since the CLM Hamiltonian $ H_\text{CL} $ is quadratic, the steady state of the probe will be Gaussian and thus, completely described by its covariances $ [\sigma_T]_{ij} \coloneqq \frac12 \langle\{ R_i, R_j \}_+\rangle  $, where $ \vec{R}^\mathsf{T} = (q_0,p_0) $ and $\{\cdot,\cdot\}_+$ denotes anticommutator. Note that the first moments vanish, (i.e., $ \langle q_0 \rangle = \langle p_0 \rangle = 0 $).

The central quantity describing the probe-sample interaction is the \textit{spectral density}
\bea \label{specd}
J(\omega) = \pi\sum_{\mu}\frac{g_\mu^2}{\omega_\mu}\delta(\omega - \omega_\mu).
\eea
Some popular profiles for $ J(\omega) $ are the Ohmic spectrum with Lorentz-Drude cutoff, i.e., $ J(\omega) = 2\gamma\omega\omega_c^2/(\omega^2 + \omega_c^2) $, which makes it particularly easy to obtain explicit analytical formulas for the steady state of the probe \cite{PhysRevA.86.012110,Correa_2017}; or the case of variable ``Ohmicity'' $ s $ and exponential cutoff \cite{PhysRevE.91.062123}, i.e., $ J_s(\omega) = \frac{\gamma\pi}{2}(\omega^s/\omega_c^{s-1})e^{-\omega/\omega_c} $. Whenever $ s > 1 $ we talk about super-Ohmic spectra, whereas the choice $ s < 1 $ corresponds to the sub-Ohmic case. The most general Ohmic spectral density ($ s = 1 $) should be of the form
\bea \label{eq:SD_shape}
J(\omega) = \gamma \omega f\left(\omega/\omega_c\right),
\eea
where $ \gamma $ is the so-called dissipation rate and $ f(x) > 0 $ is a dimensionless function that starts to decay rapidly to $ 0 $ as soon as $ x > 1 $, and that is smooth around $ x = 0 $, with $ f(0) = 1 $. The cutoff $ \omega_c $ places a cap on the frequency of the modes from the (infinite) sample that are \textit{effectively} coupled to the probe.  

Remarkably, general closed-form analytical expressions can be derived for the steady-state covariance matrix \cite{Weiss_1999}:
\begin{subequations}\label{eq:covariances} 
\bea \label{eq:q0}
&& [\sigma_T]_{11} = \frac{1}{\pi}\int_{0}^\infty d\omega \frac{J(\omega)}{|\alpha(\omega)|^2}\coth \frac{\omega}{2T}~~~\text{and} 
\\ \label{eq:p0}
&& [\sigma_T]_{22} = \frac{1}{\pi}\int_{0}^\infty d\omega \frac{\omega^2 J(\omega)}{|\alpha(\omega)|^2}\coth \frac{\omega}{2T},
\eea
\end{subequations}
while $ [\sigma_T]_{12} = [\sigma_T]_{21} = 0 $. The \textit{susceptibility} $\alpha(\omega)$ is
\bea \label{alph}
\alpha(\omega) \coloneqq \omega_0^2+\omega_R^2-\omega^2 - \chi(\omega)-\mathrm{i} J(\omega),
\eea
where $\chi(\omega) \coloneqq \frac{1}{\pi} \text{P} \int_{-\infty}^\infty d\omega'\frac{J(\omega')}{\omega'-\omega}$ is the Hilbert transform \cite{bateman1954tables2} of the spectral density, extended as an odd function for negative frequencies, i.e., $ J(\omega) \mapsto J(\omega) \Theta(\omega) - J(-\omega)\Theta(-\omega)$. Here, $\Theta(\omega)$ is the Heaviside step function, and P denotes the Cauchy principal value of the integral.

Using the definition of $ \pazocal{F}_T $ in Eq.~\eqref{eq:qfi} in combination with the formula for the Uhlmann fidelity between two single-mode Gaussian states in terms of their $ 2 \times 2 $ covariance matrices $ \sigma $ and $ \sigma' $ \cite{Scutaru_1998}
\begin{align}\label{eq:fidelity_covariance}
    \mathbb{F}(\sigma,\sigma') &= \frac{2}{\sqrt{\mathsf{\Lambda}+\mathsf{\Delta}}-\sqrt{\mathsf{\Lambda}}}, \; \text{with} \\
    \mathsf{\Lambda} &\coloneqq (4\det{\sigma}-1)(4\det{\sigma'}-1) \nonumber \\
    \mathsf{\Delta} &\coloneqq 4\det({\sigma+\sigma'}) \nonumber,
\end{align}
and Eqs.~\eqref{eq:q0} and \eqref{alph}, we can rigorously prove the following asymptotic behaviors for the QFI under the generic Ohmic spectral density of Eq.~\eqref{eq:SD_shape}:
\begin{subequations}\label{eq:scalings}
\begin{align} 
\pazocal{F}(T)&\sim T^2 \qquad~~~~ \text{for $ \omega_0 > 0 $ and $ T\ll\omega_0 $.}\label{plus2} \\
\pazocal{F}(T)&\sim 1/T^2\qquad \text{for $ \omega_0 = 0 $ and $ T \ll \omega_c $.}\label{minus2}   
\end{align} 
\end{subequations}

This is our second main result. While all details are deferred to Appendix \ref{app:Ohmic_CL}, here we illustrate Eqs.~\eqref{eq:scalings} in Fig.~\ref{fig:scaling_QLE}. Note that, in contrast with the results for gapped lattices, the low--$ T $ scaling of $ \pazocal{F}_T $ is not exponential, but power-law-like. As already advanced, the scaling \eqref{plus2} had been established by us for exponential and Lorentzian cutoff functions \cite{Correa_2017}. Interestingly, the same $T^2$ scaling was recently reported also in a gapless fermionic tight-binding chain in 1D \cite{hofer2017fundamental}. 

Note as well that the best-case relative error $ \frac{\delta T}{T} \sim \frac{1}{T\sqrt{M \pazocal{F}_T}} $ \textit{diverges} \cite{Correa_2017} as $ T\rightarrow 0 $ in the case of a Brownian thermometer with $ \omega_0 > 0 $. However, in Fig.~\ref{fig:scaling_QLE}(b) we see that it may be kept \textit{constant} over a wide range of arbitrarily low temperatures, by choosing $ \omega_0 \rightarrow 0 $. The idea of ``freezing'' $ \frac{\delta T}{T} $ by means of reducing the energy gap of the probe alongside the temperature is intuitive when thinking of fully thermalized finite-dimensional systems \cite{Paris_2016}. However, such direct temperature-dependent tuning seems very artificial. 

Luckily, however, the sensitivity of a probe with a \textit{finite} and fixed energy gap may be substantially increased at arbitrarily low temperatures by driving it periodically \cite{mukherjee2017high}. This would \textit{open} dissipative decay channels at frequencies corresponding to the ``Floquet harmonics'' of the dynamically controlled system \cite{alicki2014quantum}. In particular, very low-frequency harmonics can become the dominant contribution to the total $ \pazocal{F}_T $ under a suitable driving protocol, thus endowing the system with large thermal sensitivity at low temperatures. Whether or not this simple picture continues to hold beyond the key underlying assumption of weak probe-sample coupling remains an interesting open problem with obvious practical implications.

We have thus proven in full generality that any thermometric scheme well approximated by a harmonic probe undergoing Ohmic dissipation would display a distinct power-law like scaling in its low-$ T $ thermal sensitivity. At the same time, the Caldeira-Leggett model is arguably a good first approximation to many experimental situations of interest in quantum optics, NMR physics, and solid-state physics \cite{caldeira1983path, Weiss_1999, bp}. For instance, it can describe well the interaction of a nanomechanical oscillator with the radiation field \cite{Groblacher_2009, Groblacher_2009exper} or the dynamics of an impurity immersed in a Bose-Einstein condensate \cite{lampo2017bose}.


Finally, it is worth noting that the exponential scaling $ \pazocal{F}_T\sim e^{-\omega_0/T} $ that one would expect from a probe at thermal equilibrium \cite{Correa_2015} cannot be recovered from the exact treatment. Even if the Gibbs state is the stationary point of the commonly-used weak coupling Gorini-Kossakowski-Lindblad-Sudarshan quantum master equation \cite{bp}, the underpinning Born-Markov approximation breaks down unless the dissipation rate $ \gamma $ goes to zero at least linearly with $ T $. Therefore, for any \textit{finite} probe-sample coupling, there will always be a temperature below which the actual state of the probe deviates significantly from strict thermalization \cite{Nieuwenhuizen_2002, Allahverdyan_2012}.

\begin{figure}[t]
\centering
\includegraphics[width=0.45\textwidth]{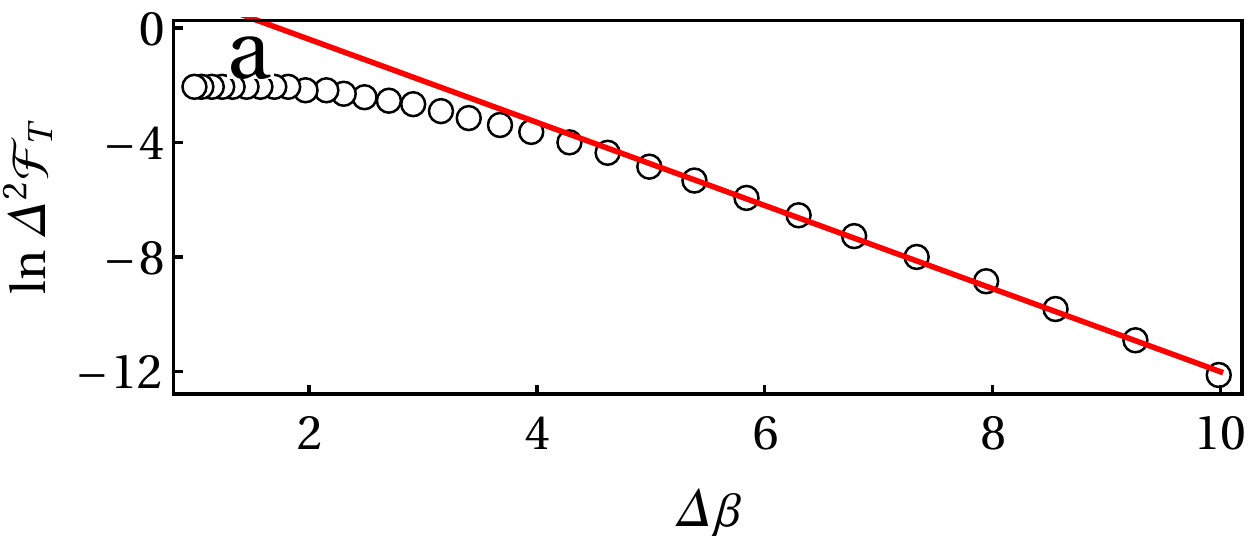}
\includegraphics[width=0.45\textwidth]{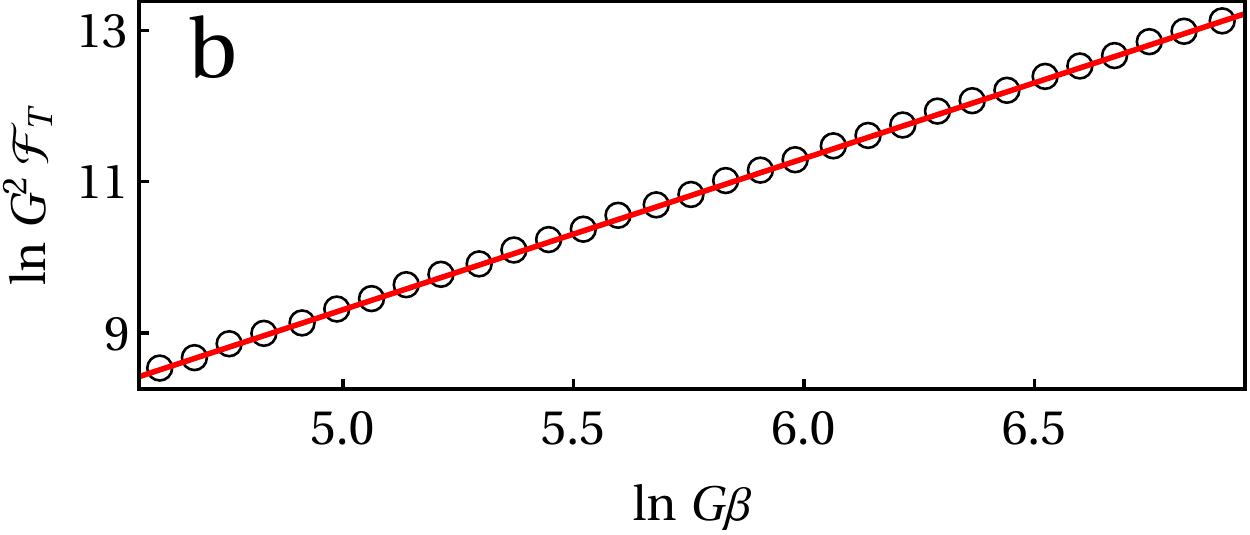}
\caption{(color online) \textbf{(a)} Log plot of the QFI as a function of the inverse temperature $ \beta $ (open circles) for a gapped translationally-invariant harmonic chain with gap $ \Delta = 0.01 $. The exponential asymptotic scaling $ \pazocal{F}_T\sim e^{-\Delta/T}$ has been superimposed in red. In this case $ N = 100 $ and the interactions have been chosen as $ G_n = G/n^t $, with $ G = 1$ and $ t = 2.5 $. For these parameters, the corresponding bare oscillator frequency is $ \Omega^2 \simeq 1.73425 $. \textbf{(b)} Log-log plot of the QFI versus the inverse temperature $ \beta $. All parameters are the same as in (a) except for $ \Omega^2 \simeq 1.73435 $, chosen so that the TIHC is gapless. The power-law-like divergence $ \pazocal{F}_T \sim 1/T^2 $ (or constant relative error $ \sqrt{M}\delta T/T $) of Eq.~\eqref{minus2} has been plotted in red.
}
\label{fig:gapped_gapless_TIHC}
\end{figure}

\section{Discussion}

In what follows, we will search for a physical intuition connecting our two main results. Our focus will be on 1D harmonic chains, which are frequently used to model crystal lattices in solid-state physics \cite{AshcroftMermin}. We will identify the key factor whereby the performance of local thermometry on a harmonic chain scales exponentially or as a power law, when $ T\rightarrow 0 $; namely, whether or not its spectral gap vanishes.

\subsection{Local thermometry on gapped and gapless harmonic chains}\label{sec:gapped_gapless_TIHC}

Our workhorse will be a 1D chain of $ 2N + 1 $ identical harmonic oscillators of frequency $ \Omega $, i.e., a TIHC, prepared at temperature $ T $ (the reason for choosing $ 2N + 1 $ nodes will become clear below). Let its Hamiltonian be
\begin{align} \label{eq:TIHC_hamiltonian}
H_\text{C} = \sum_{i=1}^{2N+1} \left( \frac{P_i^2}{2} + \frac{\Omega^2}{2} Q_i^2 \right)& + \frac{1}{2}\sum_{i \neq k}G_{ik}Q_i Q_k 
\nonumber\\
&= \frac{1}{2} \vec{P}^\mathsf{T} \vec{P} + \frac{1}{2} \vec{Q}^\mathsf{T}\,\mathsf{V}_C\,\vec{Q},
\end{align}
where $ (Q_i,P_i) $ are the quadratures of each oscillator (collected in the $ 2N + 1 $--dimensional vectors $ \vec{Q} $ and $ \vec{P} $), and $ G_{ik} $ are their mutual couplings. We will assume that these depend only on the ``distance'' between nodes, i.e., $ G_{ij} = G_{\vert i-j \vert} $ and will impose periodic boundary conditions $ G_n = G_{2N+1-n} $ for $ 1 \leq n \leq 2N $, which results in a \textit{circulant} \cite{Gray_2006} interaction matrix $ \mathsf{V}_C $, the first row of which is $ (\Omega^2, G_1, \cdots, G_N, G_N, G_{N-1}, \cdots, G_1) $. The eigenvalues of the interaction matrix (i.e., the squared normal mode frequencies of the system) are
\bea \label{eigsC2}
\Omega_a^2 = \Omega^2 + 2\sum_{k=1}^N G_k \cos{\frac{2\pi k a}{2N+1}}, \; \text{for} \; a = 0, ..., 2N,
\eea
where we notice that the frequencies $ \Omega_{N+1}, \cdots, \Omega_{2N} $ coincide with $  \Omega_N,\cdots,\Omega_1 $, respectively.

In the next section, we will comment further on the choice of the inter-node couplings. For now, we will just assume that their strength decreases with the distance, i.e., $ G_i > G_j > 0 $ for $ i < j $. In this case, the fundamental mode of the system has squared frequency $ \Delta^2 $ such that 
\begin{equation}
\Delta^2 = \Omega^2_N = \Omega^2 + 2 \sum_{k=1}^N G_k \cos{\frac{2\pi k N}{2N +1}}.
\label{eq:gap_harmonic}    
\end{equation}
Therefore, for the system's spectrum to be bounded from below, one must have
\begin{equation}\label{eq:positivity_TIHC}
\Omega^2 \geq - 2 \sum_{k=1}^N G_k \cos{\frac{2\pi k N}{2N +1}}.
\end{equation}
The strict inequality gives rise to a \textit{gapped} TIHC, whereas its saturation yields a \textit{gapless} system. 

We will now evaluate the QFI of a single node of the chain (say node $ \#1 $, as they are all equivalent) both in the gapped and the gapless case. To that end, we need to compute the elements of its reduced covariance matrix, i.e. $ [\sigma_T]_{11} = \langle Q_1^2 \rangle $ and $ [\sigma_T]_{22} = \langle P_1^2 \rangle $ ($ [\sigma_T]_{12} = [\sigma_T]_{21} = \frac12\langle\{Q_1,P_1\}_+\rangle = 0 $). Letting $ \vec{q}^C = \mathsf{O}_C \vec{Q} $ be the normal mode quadratures, these write as
\begin{subequations}
\begin{align}
[\sigma_T]_{11} &= \sum_{j=1}^{2N+1} [\mathsf{O}_C^\mathsf{T}]_{1j}^2 \frac{1}{2\Omega_{j-1}}\coth{\frac{\Omega_{j-1}}{2T}}\\
[\sigma_T]_{22} &= \sum_{j=1}^{2N+1} [\mathsf{O}_C^{\mathsf{T}}]_{1j}^2\frac{\Omega_{j-1}}{2}\coth{\frac{\Omega_{j-1}}{2T}}.
\end{align}
\label{eq:single_node_TIHC}
\end{subequations}

From Eqs.~\eqref{eq:qfi}, \eqref{eq:fidelity_covariance}, and \eqref{eq:single_node_TIHC}, one can calculate the corresponding local QFI. In Fig.~\ref{fig:gapped_gapless_TIHC}, we work out a $ 100 $--node example: As it can be seen, when the chain is gapped, the low--$ T $ sensitivity decays exponentially, as expected [see Fig.~\ref{fig:gapped_gapless_TIHC}(a)]. Note that the result in Sec.~\ref{sec:exponential_inefficiency} does not directly apply here as the interactions (and correlations) do not decay exponentially. This shows that the exponential inefficiency of local thermometry holds for a wider class of gapped systems than those with finite-range or exponentially decaying interactions. However, when the system is in the vicinity of its quantum critical point \cite{sachdev}, namely, when it is tuned to be gapless, it exhibits a power-law-like divergence of the type $ \pazocal{F}_T \sim 1/T^2 $ [see Fig.~\ref{fig:gapped_gapless_TIHC}(b)]. In view of the results about Ohmic Brownian thermometers, this observation could be the ``smoking gun'' of a deeper connection between local thermometry on many-body lattices and open system-models. We will now follow this lead by characterizing the open-system-like analogue of a single node within gapped and gapless TIHCs.

\begin{figure*}
   \includegraphics[scale=0.8]{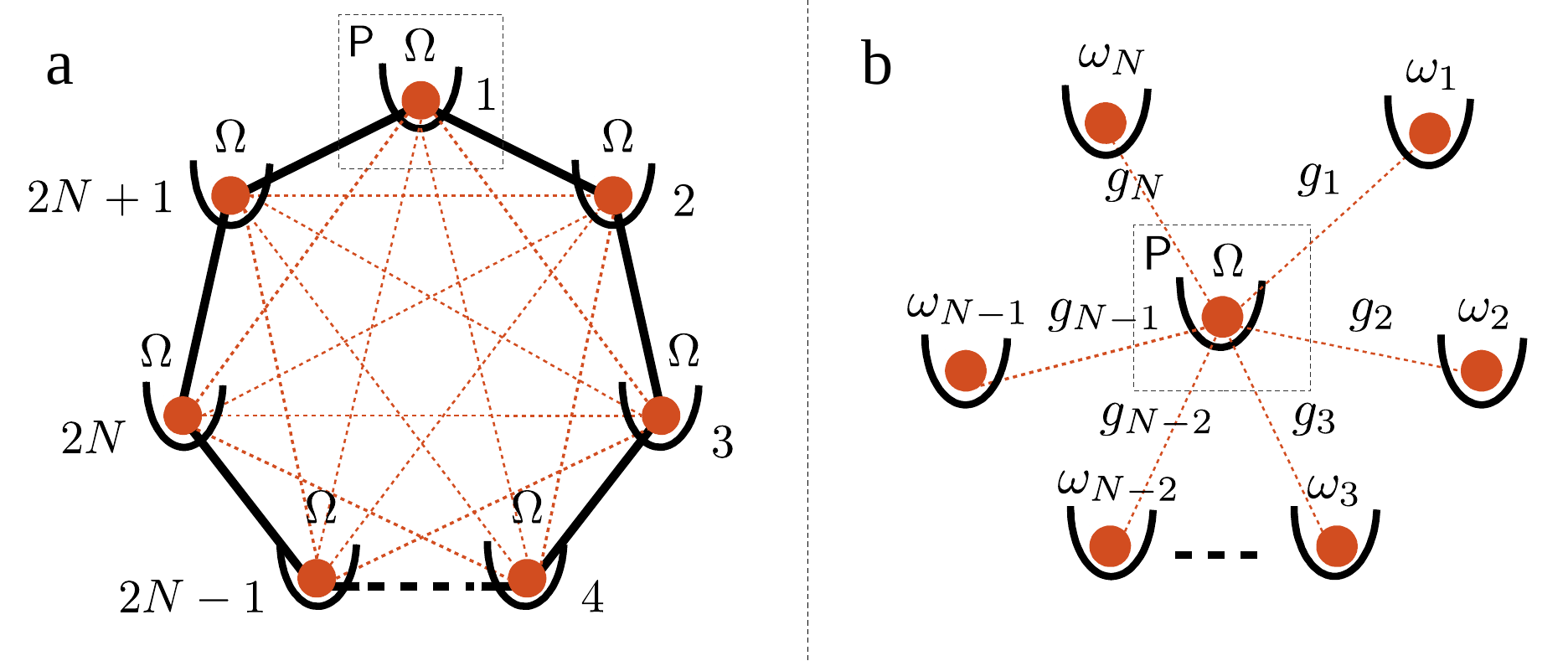}
    \caption{(color online) \textbf{(a)} Sketch of a translationally invariant Harmonic chain with periodic boundary conditions and $ 2N + 1 $ nodes of frequency $ \Omega $. Node $ \#1 $ appears highlighted as the probe $ P $. This corresponds to Eq.~\eqref{eq:TIHC_hamiltonian}. \textbf{(b)} Equivalent ``star'' system, described by Eq.~\eqref{eq:transformed_TICH}, where node $ \#1 $ couples to $ N $ of the two-fold degenerate normal modes of the inaccessible part of the chain, with frequencies $ \omega_i $.}
\label{fig:TIHC2CL}
\end{figure*}

\subsection{Mapping a translationally invariant harmonic chain into a ``star'' model and back}\label{sec:TIHC2CL_and_back}

\subsubsection{From a harmonic chain to a star model}

We shall start by splitting the Hamiltonian $ H_C $ of the TIHC in Eq.~\eqref{eq:TIHC_hamiltonian} into its accessible (i.e., node $ \# 1 $) and inaccessible parts (i.e., all other nodes), and the interactions between the two. That is,
\begin{align} \label{VC1}
H_\text{C} &= \frac12 (P_1^2 + \Omega^2 Q_1^2) + Q_1\sum_{i > 1} G_{1i} Q_i
\nonumber\\
&+ \frac{1}{2} \sum_{i>1} (P_i^2 + \Omega^2 Q_i^2) + \frac{1}{2} \!  \sum_{\substack{i\neq k \\ i,k>1}} \! Q_i [\mathsf{V}_{1|C}]_{ik}Q_k,
\end{align}
where the $ 2N \times 2N $ matrix $ \mathsf{V}_{1|C} $ results from removing the first row and column from $ \mathsf{V}_C $. Note that $ \mathsf{V}_{1|C} $ is thus not circulant ($ G_{2N-1} = G_2 \neq G_1 $) but a symmetric Toeplitz matrix \cite{Gray_2006}.

Let us denote $ \vec{Q}_{1|C}^\mathsf{T} \coloneqq (Q_2, \cdots, Q_{2N+1}) $. Provided that the real orthogonal matrix $ \mathsf{O}_{1|C} $ diagonalizes $ \mathsf{V}_{1|C} $ (i.e., $ \mathsf{O}_{1|C}\mathsf{V}_{1|C}\mathsf{O}_{1|C}^\mathsf{T} = \diag\{{\omega_1^2,\cdots,\omega_{2N}^2}\}$), we define the \textit{sample} degrees of freedom from the normal-mode coordinates of the inaccessible nodes $ \vec{q}^{1|C}\coloneqq\mathsf{O}_{1|C}\vec{Q}_{1|C} $. Eq.~\eqref{VC1} thus rewrites as
\begin{align}
H_C = \frac12(P_1^2 + \Omega^2 Q_1^2) &+ Q_1 \sum_{i>1} g_i q^{1|C}_i \nonumber\\
&+ \frac12\sum_{i>1} \big[(p_i^{1|C})^2 + \omega_i^2 (q^{1|C}_i)^2\big].
\label{eq:transformed_TICH}
\end{align}
The transition between the chain-like model of Eq.~\eqref{VC1} and the star-like configuration of Eq.~\eqref{eq:transformed_TICH} is depicted in Fig.~\ref{fig:TIHC2CL}. 

Note that the transformed coupling constants are given by $ g_i = [\mathsf{O}_{1|C}]_{ij}G_{1j} $. Due to the existing symmetries, the probe interacts only with \textit{half} of the sample modes. Therefore, we shall keep only the $ N $ relevant ones and define the effective spectral density as $ J(\omega) = \pi\sum_{i=1}^N (g_i^2/\omega_i)\delta(\omega-\omega_i) $, which will be the central object of interest in what follows. 

\begin{figure}[b]
  \includegraphics[width=\columnwidth]{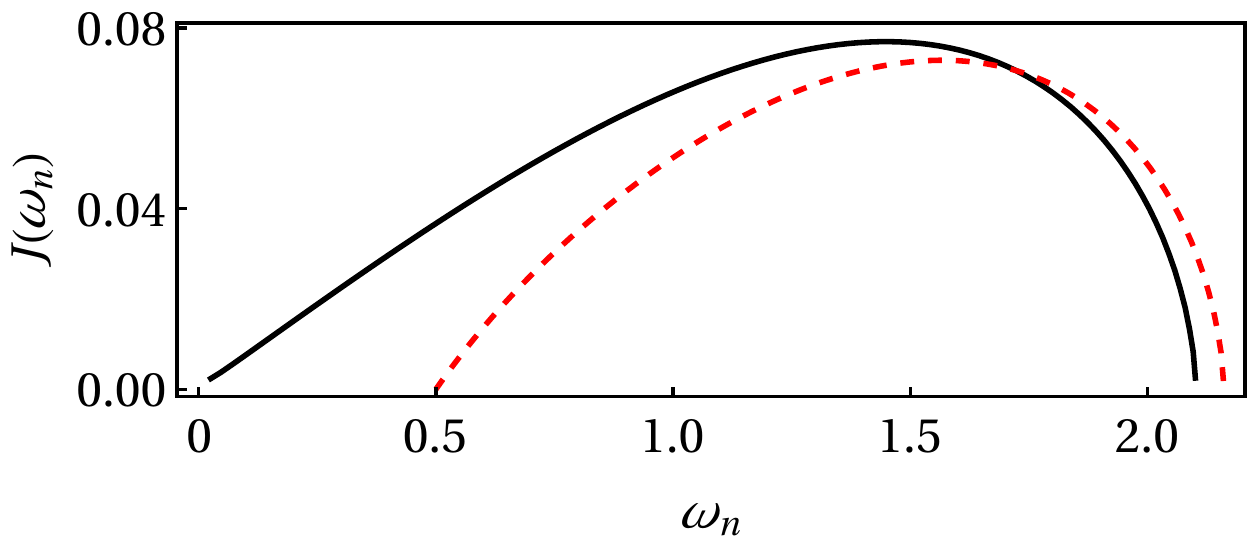}
    \caption{(color online) Effective spectral densities after partitioning an $ N = 100 $ TIHC as ``node $ \# 1 $'' versus the rest. The interactions $ G_n $ are the same as in Fig.~\ref{fig:gapped_gapless_TIHC}. We plot both the gapless \textbf{(solid black)} and the gapped case \textbf{(dashed red)}, with gap $ \Delta = 0.5 $ (i.e., $\Omega^2\simeq 1.98425 $). Note that in both situations $ J(\omega) $ shows an approximately linear growth for low $ \omega_n $ and features a cutoff. In the gapped case, however, the lowest sample mode coupled to the probe has a finite frequency and hence, the resulting spectral density is \textit{not} Ohmic.}
\label{fig:TIHC_spectral_dens}
\end{figure}

As illustrated in Fig.~\ref{fig:TIHC_spectral_dens}, whenever $ G_n $ decays as $ n^{-1} $ or faster, the spectral density is approximately linear around its minimal frequency, which is non-zero. Hence, gapped TIHCs are not capable of reproducing the canonical Ohmic form of Eq.~\eqref{eq:SD_shape} in their residual spectra. In any case, they are endowed with a high-frequency cutoff. 

In the limit of large $ N $, the chain is gapless for $ \Omega^2 \simeq 2\sum_{n=1}^{\infty}(-1)^{n-1} G_n $ as follows from the saturation of Eq.~\eqref{eq:positivity_TIHC} (see Appendix \ref{app:chain_gap} for a discussion on the error bars of this approximation), and the effective spectral density becomes truly Ohmic (see Fig.~\ref{fig:TIHC_spectral_dens}). We note that a closing gap also implies $ \Omega^2 = \omega_R^2 $, where $\omega_R$ is the Caldeira-Leggett renormalization frequency.

These facts leave us with the following picture: Whenever the internal couplings in a gapless TIHC decay at least as fast as the inverse of the distance between the nodes, the interaction of every node with the rest of the chain is described by an Ohmic Brownian motion model, in which the probe has vanishing bare frequency $ \omega_0 = 0 $. This means that Eq.~\eqref{minus2} can be directly applied when the temperature of cold TIHC is to be estimated by measuring a single node. In turn, this is consistent with our observations in Fig.~\ref{fig:gapped_gapless_TIHC}(b). 

Limiting ourselves to short-range interacting nodes is mostly a technical requirement that allows us to circumvent potential problems derived from the super-extensive scaling of the energy \cite{PhysRevLett.87.030601}. Nonetheless, we have numerically explored a large range of TIHCs: In addition to the standard choices of algebraic (i.e., $ G_n \propto n^{-t}$ for $t > 1 $, as in Fig.~\ref{fig:gapped_gapless_TIHC}) and exponential ($ G_n \propto e^{- c n} $ for $ c > 0 $) interactions, we have run tests using ordered lists of random numbers as coupling constants. In all cases, the results were qualitatively the same, which makes us confident that they hold in general.	

\subsubsection{From a star model to a harmonic chain}\label{sec:star_to_TICH}

Let us consider the reverse problem, i.e., finding a TIHC that corresponds to a given (discretized) CLM. At the most basic level, one wants to ensure that there exists a TIHC with the same set of normal modes as the linear open-system at hand. However, the probe in the CLM will not correspond, in general, to one of the nodes of its associated TIHC. Rather, it will be \textit{delocalized} over some (or all) of its nodes. This is due to the fact that the canonical transformations diagonalizing both systems are generally different. It is thus interesting to determine how does local thermometry on a CLM look from the perspective of its TIHC analogue.

\begin{figure}[t]
   \includegraphics[width=\columnwidth]{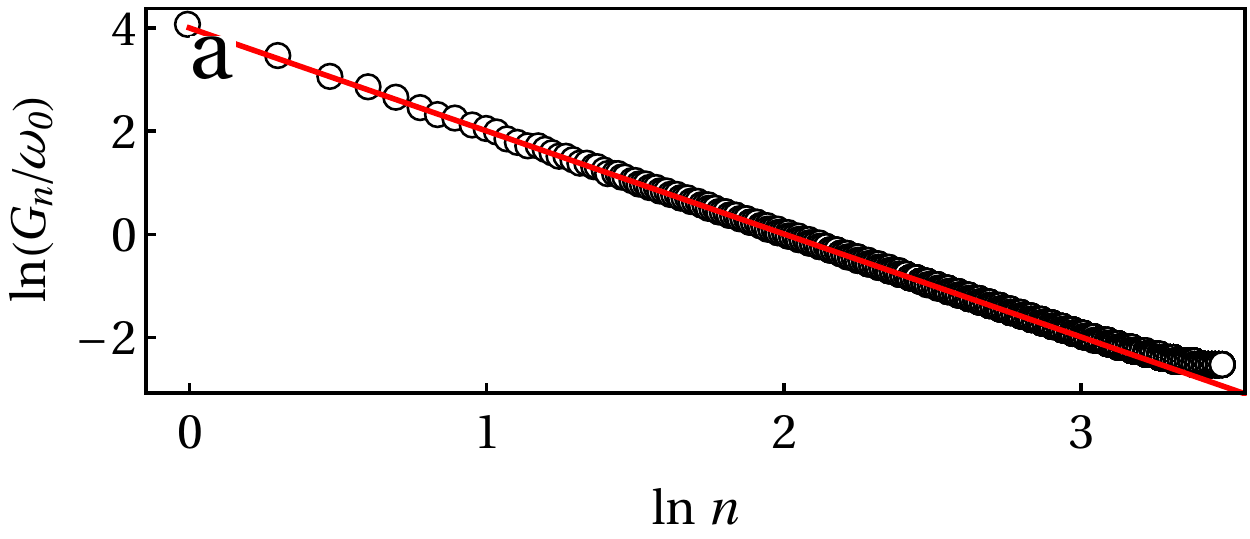}
   \includegraphics[width=\columnwidth]{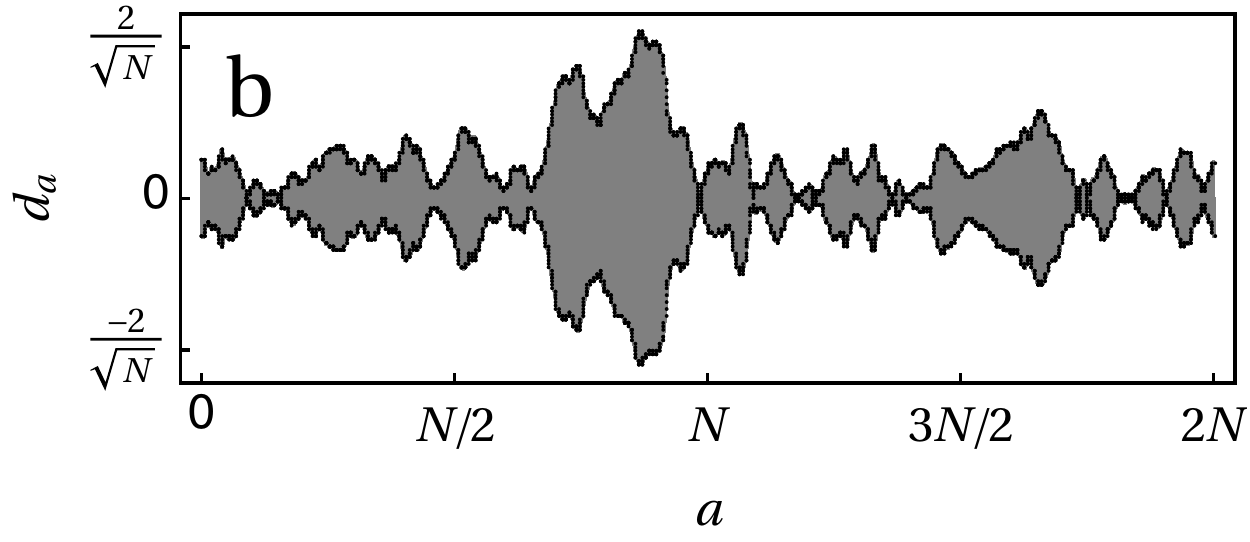}
    \caption{(color online) \textbf{(a)} Log-log plot of the coupling constants $ G_n $ (open circles) corresponding to a discretized Caldeira-Leggett model with Ohmic spectral density and Lorentz-Drude cutoff, as a function of the inter-node distance $ n $. In particular, the spectral density is characterized by $ \gamma = 0.1 $ and $ \omega_c = 2 $. A total of $ N = 2000 $ frequencies were uniformly picked within the interval $ \omega_n\in(0,\omega_{\max}) $, with $ \omega_{\max} = 100 $. The frequency of the probe was chosen as $ \omega_0 = 0.2 $. A linear fit $ G_n \sim n^{-c} $ with $ c \simeq 1.99469 $ (red) has been added for comparison (see text for further details). \textbf{(b)} Coefficients of the expansion $ q_0 = \sum_a d_a Q_a $ of the position of the probe in the CLM as a linear combination of the positions of the oscillators in the corresponding TIHC. All parameters are the same as in (a).}
\label{fig:CL2TIHC}
\end{figure}

Let us recall from the preceding sections that all but one (namely, the largest, $ \Omega_0 $) of the normal modes of a $ 2N +1 $--node TIHC are doubly degenerate. We can define the $N+1$--dimensional vectors $ \vec{\Omega}^\mathsf{T} \coloneqq (\Omega_0^2,\Omega_1^2,\cdots,\Omega_N^2) $ and $ \vec{G}^{\mathsf{T}} \coloneqq (\Omega^2,2G_1,\cdots,2G_N) $, containing the non-repeated normal-mode squared frequencies, and the squared ``physical'' frequency and interaction strengths, respectively. Hence, we may rewrite Eq.~\eqref{eigsC2} in compact form as $ \vec{\Omega} = \mathsf{A} \vec{G} $, where $[\mathsf{A}]_{jk} = \cos{(\frac{2\pi jk}{2N+1})}$, for $ j,k = 0,\cdots,N $. 

Proceeding from the other end, we may calculate the normal mode frequencies of the $ N + 1 $--particle CLM under consideration and arrange them in decreasing order in the vector $ \vec{\omega}^\mathsf{T} \coloneqq (\omega_0^2, \cdots, \omega_N^2) $. All we have to do is to invert the above relation, i.e., $ \vec{G} = \mathsf{A}^{-1}\vec{\omega} $, so that the resulting $ \vec{G} $ fully characterizes the TIHC matching our CLM. 

In order to put the above into practice, we need to discretize a CLM. We start by setting a cap on the sample frequencies $ \omega_{\max} > \omega_c $ and distributing our frequencies uniformly over the allowed range. In order to obtain the coupling $ g_n $ for any given $ \omega_n $, one may use the relation
\begin{equation}\label{eq:discrete_couplings_CL}
g_n^2 = \frac{\omega_n}{\pi} \int_{\Delta\omega} d\omega J(\omega),
\end{equation}
which follows from Eq.~\eqref{specd} whenever the frequency interval $ \Delta\omega $ is chosen around $ \omega_n $ so that neither of the neighbouring sample node frequencies $ \omega_{n\pm 1} $ are contained in it. In order to make sure that the discrete model represents its continuous counterpart faithfully, it is sufficient to require
\begin{equation}\label{eq:CL_discretization_condition}
    \omega_R^2 = \frac{2}{\pi}\int_0^\infty d\omega \frac{J(\omega)}{\omega} \simeq \sum_{n=1}^N \frac{g_n^2}{\omega_n^2}.
\end{equation}
We discuss this point in Appendix \ref{app:correct_limit}, for the Ohmic spectral density with Lorentz-Drude cutoff (introduced above), and show that a choice of parameters such that $ N \gg\omega_{\max}/\omega_c \gg 1 $ guarantees a good agreement.

In Fig.~\ref{fig:CL2TIHC}(a) we illustrate this calculation for $ N = 2000 $, $ \omega_0 = 0.2 $, $ \gamma = 0.1 $, $ \omega_c = 2 $, and $ \omega_{\max}=100 $, for which $ \sum_n g_n^2/\omega_n \simeq 0.195853 \simeq \omega_R^2 = 0.2 $. The nodes of the TIHC corresponding to these parameters can be found to have frequency $ \Omega \simeq 57.7278 $, while the couplings decay as a power law $ G_n \propto n^{-c} $ with almost constant $ c $. Until around $ n = 500 $, $ c \simeq 2 $. For $ 500 \lesssim n \lesssim 1000 $, $ c $ becomes slightly smaller (i.e., $ c \approx 1.77 $), although it remains approximately constant. Finally, for $ n \gtrsim 1000 $, the decay of interactions becomes slower and non-power-law-like. Considering different values of $ N $, we find the above change of behavior to occur around $ N/2 $. We are thus dealing with a finite-size effect which does not appear in the thermodynamic limit. In other words, in the thermodynamic limit, the TIHC corresponding to an Ohmic CLM \cite{Correa_2017} becomes a gapless chain with interactions decaying as the square of the distance. 

We know from Eq.~\eqref{plus2} that the QFI scales as $ \pazocal{F}_T \sim T^2 $ at low temperatures in this model. However, we also know that optimal local thermometry on a gapless TIHC should give rise to the diverging low--$ T $ behavior $ \pazocal{F}_T\sim 1/T^2 $. The reason for this discrepancy is that, a local measurement on the central oscillator of the CLM does not map into a local measurement on one node of the corresponding TIHC---it rather maps into a complex measurement which turns out to be sub-optimal, in spite of spreading over the whole chain. We will conclude this discussion showing that this is indeed the case.

Now, let the Caldeira-Leggett Hamiltonian in Eq.~\eqref{eq:Caldeira_Leggett} be written as $ H_{CL} = \frac12(\vec{P}^{CL})^\mathsf{T}\vec{P}^{CL} + \frac12 (\vec{Q}^{CL})^\mathsf{T}\mathsf{V}_{CL}\vec{Q}^{CL} $, and its $ N + 1 $ normal-mode coordinates, as $ \vec{q}^{CL} = \mathsf{O}_{CL} \vec{Q}^{CL} $. On the other hand, let $ \vec{q}^C $ contain the $ 2N + 1 $ TIHC normal-mode quadratures $ \vec{q}^C = \mathsf{O}_{C} \vec{Q}^C $. We assume that these are ordered in such a way that the first $ N + 1 $ elements of $ \vec{q}^C $ correspond to the non-degenerate frequencies $ \{ \Omega_0, \Omega_1,\cdots,\Omega_N\} $. One can thus connect the original CL coordinates with those of the TIHC via $ \vec{Q}^{CL} = (\mathsf{O}_{CL}^\mathsf{T}\oplus\mathbbm{1}_N)\mathsf{O}_C\vec{Q}^C $. In particular, the matrix elements $ d_a \coloneqq [(\mathsf{O}_{CL}^\mathsf{T}\oplus\mathbbm{1}_N)\mathsf{O}_C]_{1a} $ are the coefficients in the expansion of the position of the probe in terms of the positions of the oscillators in the chain, i.e. $ q_0 = \sum_a d_a Q_a $. 

We plot the coefficients $ d_a $ in Fig.~\ref{fig:CL2TIHC}(b) for the same parameters of Fig.~\ref{fig:CL2TIHC}(a). We can see that $ q_0 $ spreads all over the chain and hence, local manipulations of the probe on the CLM map into complex global measurements on the corresponding TIHC. Notice however, that the resulting low-temperature scaling of the QFI (i.e., $ \pazocal{F}_T\sim T^2 $) is far worse that what could be achieved by interrogating locally a single node of the chain [cf. Eq.~\eqref{minus2}]. We thus see how local thermometry on a simple linear system can turn into a surprisingly rich problem.

\section{Conclusions}\label{sec:discussion}

In this paper we have focused on local thermometry on quantum many-body systems, deep into the low temperature regime. First, we proved that the accuracy of local thermometry is exponentially suppressed for any \textit{gapped}, translationally invariant, non-critical and short-range-interacting lattice system. This result is very general and applies to locally-finite as well as harmonic many-body systems.

Furthermore, in order to explore the \textit{gapless} regime, we adopted an open-system approach, and established that thermometry on a harmonic probe coupled to an Ohmic sample is characterized by a distinctive power-law-like low-temperature scaling. Namely, Brownian particles with finite bare frequency (i.e., $\omega_0 \neq 0$) can sense the temperature of a much larger equilibrium sample with a precision scaling as $ \sim T^2 $ for $ T\rightarrow 0 $. On the contrary, a probe with $ \omega_0 \rightarrow 0 $ displays a remarkable diverging low--$ T $ scaling of $ \sim 1/T^2 $ for a wide range, extending to arbitrarily low temperatures. Most importantly, we showed that the decisive factor when switching between exponential and sub-exponentially inefficient quantum thermometry is whether or not the energy spectrum of the global many-body system exhibits a finite gap.

In order to gain further insights into the problem, we studied a simple 1D chain of identical harmonic oscillators with arbitrary interactions and periodic boundary conditions. Our main finding was that the open-system formulation of local thermometry on a single node of a gapped instance of such chain gives rise to a non-standard dissipative model in which the probe is effectively decoupled from the lower-frequency modes of the sample. For the probe to be able to interact with all the sample modes in the open system description, the 1D chain must be initially gapless. It is intuitively clear that, at sufficiently cold temperatures, those neglected low-frequency modes become dominant. In turn, this explains the exponential suppression of thermal sensitivity in gapped systems. 

Remarkably, we demonstrated that local thermometry on a node of a translationally invariant gapless harmonic chain, with interactions of arbitrary strength and range (provided they decay at least as fast as the inverse of the distance between the nodes), can be mapped to the problem of estimating the temperature of an \textit{Ohmic sample} with a harmonic probe at $\omega_0 = 0$. 

Additionally, we discussed how to discretize and map a continuous open system model of the Caldeira-Leggett type into a translationally invariant harmonic chain. We were thus able to show that the resulting chain exhibits interactions that decay quadratically with the inter-node distance. Finally, we illustrated how, in spite of the necessary existence of a formal open-system-to-chain mapping, the individual Brownian probe needs not correspond to a single node in the chain. Instead, local manipulations of the probe generally look like complex global manipulations on the corresponding chain. 

Our results thus shed light on the technologically relevant problem of sensing ultra-cold temperatures from various different angles. Even though we make fundamental statements about the ultimate low-temperature limitations on the precision of temperature measurements, our results have also clear practical implications. Note for instance that the scaling $ \pazocal{F}_T \sim 1/T^2 $ of the sensitivity of a Brownian particle with vanishing bare frequency implies that the relative error $ \delta T / T = (T\sqrt{\pazocal{F}_T})^{-1} $ can be kept \textit{constant} for arbitrarily low $ T $, by simply tuning the trapping frequency of the thermometer to be sufficiently low. This observation is intimately connected with a recent proposal on low-$ T $ thermometry exploiting dynamical control \cite{mukherjee2017high}.

Let us also point out that, since the state of the locally measured probes is the reduction of a global thermal state, we might invoke typicality arguments to extrapolate our results to a coherent superposition of eigenstates of the global Hamiltonian in a small energy window \cite{Popescu_2006, Goldstein_2006, Muller_2015}, or sometimes even a single such eigenstate \cite{Rigol_2008}. Note, however, that care must be taken when applying such typicality results in the low-$T$ limit \cite{Reimann_2012, Gogolin_2016}.

Our setting is also well suited for tackling other interesting situations, such as local thermometry on gapped \textit{long-range-interacting} systems. In fact, the chain-to-open-system mapping could potentially be exploited to solve such problem \textit{exactly} once the corresponding effective spectral density has been worked out. Moreover, since most second-order classical phase transitions take place at rather low temperatures, our methods can be useful also in the problem of local thermometry in critical systems. There, the thermal sensitivity is expected to be increased due to the presence of long-range correlations \cite{Mehboudi_2016, Marzolino_2013, Zanardi_2008}. These issues are worthy of further investigation and will be considered elsewhere.

\section*{Note added}
During the completion of this manuscript, we became aware of the closely related work by P. P. Hofer \textit{et al.} \cite{hofer2017fundamental}. In it, the authors argue that gapless spectra allow for sub-exponential low--$ T $ scaling of $ \pazocal{F}_T $ and work out several examples, including local thermometry in a 1D tight-binding fermionic chain, which also leads to the scaling $ \pazocal{F}_T \sim T^2 $. 

\section{Acknowledgements}

We thank P. P. Hofer, J. B. Brask, and N. Brunner for their useful feedback on this manuscript. We gratefully acknowledge funding by the Villum Fonden, the European Research Council (ERC) Starting Grant GQCOP (Grant No. 637352), and the COST Action MP1209: ``Thermodynamics in the quantum regime''.

\appendix

\section{Heat capacity of local Hamiltonians}\label{app:heat_capacity}

In this first appendix, we shall argue that the large-size and low-temperature scaling of the heat capacity of gapped translationally-invariant lattices with finite-range two-body interactions is
\bea \label{capaz}
C_N(T) \leq \pO( N e^{-\beta\Delta}).
\eea
For harmonic and free-fermion lattices in arbitrary spatial dimensions we prove this in full generality. Although the extensivity is trivial to show also for general lattices with locally finite Hilbert space dimension, the exponential temperature dependence is far more challenging to prove. However, by force of example, we expect Eq.~\eqref{capaz} to hold also in this case.

\subsection{Extensivity}

In short-range interacting, translationally-invariant lattice systems with finite local Hilbert space dimension, extensivity of the heat capacity---i.e., that $\lim\limits\nolimits_{N\to\infty}C_N/N$ exists and is finite---is a trivial consequence of the fact \cite{Araki_1969, Araki_1974} that, for the partition function of a translationally-invariant system, the limit $\lim\limits\nolimits_{N\to\infty}N^{-1}\ln Z_N$ exists and is regular. Indeed, this means that, for $N\gg1$, $\ln Z_N = \pO(N)$, and the extensivity of $C_N$ follows from the identity $C=\beta^2\partial_\beta^2\ln Z$.

More intuitively, and, most importantly, also applicable to harmonic lattices, the extensivity can be understood as follows. By representing the lattice by a graph $(V_N,E_N)$, where $V_N$ is the set of all sites serving as vertices, and the edges $E_N$ are the interacting pairs, we write the Hamiltonian as
\bea
H_N = \sum_{v\in V_N} H_v + \sum_{e\in E_N}h_e,
\eea
where $H_v$ are the on-site Hamiltonians, and $h_e$ are the interactions. Since we consider only regular lattices and interactions of finite range, there is only a finite set of edges connected to each vertex. We denote it by $E_v$ and rewrite $H_N$ as
\bea
H_N = \sum_{v\in V_N} \big(H_v + \sum_{e\in E_v} h_e/2\big) \equiv \sum_{v\in V_N} \tilde{H}_{E_v},
\eea
where $\tilde{H}_{E_v}$ ``lives'' in the joint Hilbert space of the vertices at the ends of the edges in $E_v$. Due to translational invariance, the operators $\tilde{H}_{E_v}$ and the corresponding marginal states $\rho^{(N)}_{E_v}$ (Note the dependence of the local state on the global system size), are the same for any $ v $, in spite of ``residing'' in different Hilbert spaces. Hence, for the energy of the lattice, we have
\bea 
E_N = N \tr(H_{E_1}\rho^{(N)}_{E_1}).
\eea

Finally, by the very definition of the problem, the global state converges when $N\to\infty$ (see also \cite{Araki_1969, Araki_1974}), and therefore we may formally write $\rho^{(N)}_{E_1}=\rho^{(\infty)}_{E_1} + o(1)$. And since $\rho^{(\infty)}_{E_1}$ is either a finite-component bosonic Gaussian state or a finite-dimensional positive operator of trace $1$, $\tr(H_{E_1}\rho^{(N)}_{E_1}) = e(T) + o(1)$, where $e(T)$, away from criticality, is a regular function of $T$ and is $\pO(1)$. Hence, we conclude that $E_N = N e(T) + o(N)$, which thus proves the extensivity of $C_N=dE_N/dT$. This allows us to define the specific heat $c(T) \coloneqq \lim\limits\nolimits_{N\to\infty}C_N(T)/N$, which we shall study below.

\subsection{Specific heat at low temperatures} \label{app:finite_dimensional_heat_capacity}

Let us first prove Eq.~\eqref{capaz} for harmonic and free-fermion systems. These are widely used to describe a large variety of physical objects, from quantum fields to superconductors (see, e.g., \cite{Wen_2004}), and are described by Hamiltonians that are bilinear in suitably defined bosonic or fermionic creation and annihilation operators. Due to their bi-linearity, such Hamiltonians can always be canonically decomposed as ``normal modes'', i.e., $\sum_{n=1}^N\epsilon_n a_n^\dagger a_n$, where $a^\dagger_n$ and $a_n$ are the (bosonic or fermionic) creation and annihilation operators, $\epsilon_n$ are the normal mode frequencies, and $N$ is the number of lattice sites. For convenience, we shall arrange the set $\{\epsilon_1\leq\epsilon_2\leq...\}$ in order of increasing energy. 

In both bosonic and fermionic cases, the spectral gap $\Delta$ of the whole system will be equal to $\epsilon_1$. Furthermore, since the normal modes do not interact with each other, when the global system is in a thermal state, \textit{each} of the modes is thermal. The heat capacity of a single mode is
\bea \label{cien}
C_n = \frac{(\beta\epsilon_n)^2 e^{-\beta\epsilon_n}}{(1 \mp e^{-\beta\epsilon_n})^2},
\eea
where the minus (plus) sign should be used for a bosonic (fermionic) mode. 

The low-temperature regime is defined as $T\ll\Delta$. Thus, $\beta\epsilon_n\gg 1$ for all $n$'s. On the other hand, for large $ \beta\varepsilon_n $, the function in Eq.~\eqref{cien} is decreasing. Hence, for sufficiently low temperatures ($\beta\Delta\geq 4$ would be sufficient), we have
\bea
C_N\leq N \frac{(\beta\Delta)^2 e^{-\beta\Delta}}{(1 \pm e^{-\beta\Delta})^2}.
\eea
In particular, this implies that
\bea
c(T)\leq (\beta\Delta)^2 e^{-\beta\Delta}\times\pO(1),
\eea
which proves Eq.~\eqref{capaz}. Moreover, it points to the fact that one can add a $(\beta\Delta)^b$ multiplier to Eq.~\eqref{capaz} in order to sharpen the asymptotics. Here, $b$ would be some system-dependent number. For example, in Appendix \ref{app:Ising}, we carefully calculate the specific heat for the quantum Ising model and show that, at low-temperatures, it scales as $(\beta\Delta)^{3/2} e^{-\beta\Delta}$. 

In fact, an identical analysis shows that this scaling holds for any system that can be decomposed into non-interacting parts, so that the dispersion relation is quadratic \cite{Heiniger_1966}. Furthermore, the scaling $(\beta\Delta)^{3/2} e^{-\beta\Delta}$ was demonstrated for the 1D spin-$1/2$ XYZ model \cite{Takahashi_1973}, which is \textit{not} of the free-fermion type. In general, it is ``folklore'' in solid-state physics that the specific heat in gapped systems decays as $(\beta\Delta)^b e^{-c(\beta\Delta)}$ ($c>0$) \cite{Corak_1954, Rosenberg_1963, Heiniger_1966, Chakraborty_1997, van_Dijk_1997, Wang_2001} (see also \cite{Brandao_2015} for a similar discussion).

\subsection{Low-$ T $ specific heat of gapped quantum Ising model} \label{app:Ising}

Let us illustrate the above result on the simple example of quantum Ising model in a transverse field \cite{Lieb_1961}. This is described by the Hamiltonian
\bea
H_\text{I}=\frac{J}{2}\sum_i\sigma_x^i\otimes\sigma_x^{i+1} -\frac{h}{2}\sum_i \sigma_z^i,
\eea
which, in the free-fermion representation \cite{sachdev}, takes the form
\begin{equation}
H_\text{I} = \sum\nolimits_{k=1}^{N}\epsilon_k (c_k^\dagger c_k-1/2),
\end{equation}
where $c^\dagger_k$ and $c_k$ are the creation and annihilation operators of the $k$--th fermionic mode, and
\begin{equation}
\epsilon_k = 2 \sqrt{J^2+h^2 - 2hJ  \cos\left(2\pi k/N\right)},
\end{equation}
for $ k\in\{ -\left\lfloor\frac{N}{2}\right\rfloor,\cdots,\left\lfloor\frac{N}{2}\right\rfloor-1 \} $. The smallest gap among the two-level systems and hence, the spectral gap of the total system, is $ \Delta=2|h-J| $. Incorporating this to the notation, one may rewrite the energies $ \varepsilon_k $ as
\bea
\epsilon_k = \Delta\sqrt{1 + \frac{16hJ}{\Delta^2}\sin^2\left(\frac{\pi k}{N}\right)}.
\eea
Since the spin chain has been mapped into a collection of non-interacting two-level systems, the total heat capacity is nothing but the sum of their individual heat capacities, i.e., $C_N(T) = \sum_k C(\epsilon_k,T)$, where
\bea \label{qubitheat}
C(\epsilon_k, T)=\frac{(\beta\epsilon_k)^2 e^{-\beta \epsilon_k}}{(1+e^{-\beta \epsilon_k})^2}.
\eea
Let us now define
\bea
k_m \coloneqq \left\lfloor\frac{N}{\sqrt[3]{\beta\Delta}} \right\rfloor\ll N
\eea
and write the heat capacity as
\bea
C_N = \sum_{k=-k_m}^{k_m}C(\epsilon_k,T) + \sum_{|k|>k_m} C(\epsilon_k,T).
\eea
Noticing that the second sum is upper-bounded by $(N-2k_m)C(\epsilon_{k_m},T)$, and keeping in mind that we are interested in the regime where $\beta\Delta\gg 1$, we get
\bea
C_N = \sum_{k=-k_m}^{k_m}C(\epsilon_k,T) + \pO \left(N e^{-\beta\epsilon_{k_m}}\right).
\eea
Furthermore, since $k_m/N\ll1$, for $k\leq k_m$, we have
\bea
\epsilon_k = \Delta + \frac{8\pi^2hJ}{\Delta}\frac{k^2}{N^2} + \pO\left[(k/N)^4 \right]
\eea
Further noticing that $(\beta\Delta) k_m^4/N^4=(\beta\Delta)^{-1/3}\ll 1$ and denoting $f\coloneqq\frac{8\pi^2hJ}{\Delta N^2}$, we obtain
\begin{align} \label{klir3}
C_N =(\beta\Delta)^2 e^{-\beta\Delta}\sum_{k=-k_m}^{k_m} &\left( 1 + \frac{2 f}{\Delta} k^2 + \pO\left[(k/N)^4 \right] \right)  e^{- \beta f k^2} \nonumber\\
&+ \pO \left(N e^{-\beta\epsilon_{k_m}}\right).
\end{align}

Finally, noticing that $\beta f k_m^2 = \pO \left(\sqrt[3]{\beta\Delta} \right)\gg 1$, one can write the Euler-Maclaurin formula \cite{fixt}
\begin{equation}
\sum_{k=-k_m}^{k_m}e^{-\beta f k^2} = \int_{-k_m}^{k_m}dx e^{-\beta f x^2} + \pO\left(e^{-\sqrt[3]{\beta\Delta}} \right).
\end{equation}
Splitting the integral as $\int_{-k_m}^{k_m}=\int_{-\infty}^{\infty} - \int_{k_m}^{\infty} - \int_{-\infty}^{-k_m}$ and noticing that the latter are $\pO\left( e^{-\sqrt[3]{\beta\Delta}} \right)$, we obtain that
\bea \label{eumac}
\sum_{k=-k_m}^{k_m}e^{-\beta f k^2} = \pi(\beta f)^{-1/2} + \pO\left( e^{-\sqrt[3]{\beta\Delta}} \right).
\eea
Deriving the both sides of Eq.~\eqref{eumac} with respect to $(\beta f)$ once and twice to find, respectively, $\sum_{k=-k_m}^{k_m} k^2 e^{-\beta f k^2}$ and $\sum_{k=-k_m}^{k_m} k^4 e^{-\beta f k^2}$, and substituting in Eq.~\eqref{klir3} yields
\bea \label{ising_heat_capacity}
C_N = N(\beta\Delta)^{3/2} e^{-\beta\Delta}\sqrt{\frac{\Delta^2}{8hJ}}\left[ 1+ \pO\left(\frac{1}{\beta\Delta}\right) \right],
\eea
which, we emphasize, is correct only for $N\gg1$ and $\beta\Delta\gg 1$.

\section{Getting Eq.~\eqref{eq:qfi_alt} from Eq.~\eqref{eq:qfi}}\label{app:from_qfi_to_qfialt}

Eq.~\eqref{eq:qfi_alt} amounts to 
\begin{align}
\mathbb{F}(\rho_T,\rho_{T+\delta}) &= 1 - \frac{1}{4} \pazocal{F}_T\,\delta^2 + \pazocal{O}(\delta^3)\nonumber
\\
&= 1 + \frac12 \left(\lim_{\delta\rightarrow 0}\frac{\partial^2\mathbbm{F}(\rho_T,\rho_{T+\delta})}{\partial\delta^2}\right)\delta^2 +  \pazocal{O}(\delta^3).
\label{eq:appA-1}    
\end{align}
Which holds provided that $ \mathbb{F}(\rho_T,\rho_{T+\delta}) = \pazocal{O}(\delta^2) $. 
To see that this is the case, let us introduce the operator $ x \coloneqq \frac{\partial \rho_T}{\partial T} \delta + \pazocal{O}(\delta^2) $, so that $ \rho_{T+\delta}=\rho_T + x $. The Uhlmann fidelity would thus rewrite as $ \mathbb{F}(\rho_T,\rho_{T+\delta})= \big(\tr\sqrt{\sqrt{\rho_T}\,(\rho_T +x)\sqrt{\rho_T}}\big)^2 = \big(\tr\sqrt{\rho_T^2+\sqrt{\rho_T}\,x\sqrt{\rho_T}}\big)^2 \coloneqq [\tr\,(\rho_T + y)]^2 = (1 + \tr y)$. 

Squaring the definition of this newly-introduced operator $ y $, we see that
\begin{equation} \label{eq:appA-2}
y^2 + \rho_T\,y + y\,\rho_T = \sqrt{\rho_T}\,x\sqrt{\rho_T}.
\end{equation}

Since $ x = \pazocal{O}(\delta) $, it is also clear that $ y = \pazocal{O}(\delta) $. We can now multiply from left and right by $ (\sqrt{\rho_T})^{-1} $, which yields
\begin{equation}
(\sqrt{\rho_T})^{-1} y \sqrt{\rho_T} + \sqrt{\rho_T} y (\sqrt{\rho_T})^{-1} = \frac{\partial\rho_T}{\partial T}\delta + \pazocal{O}(\delta^2).
\label{eq:appA-3}    
\end{equation}
Note that the \textit{invertibility} is not an issue here, even if $ \rho_T $ is not full rank, since all $ \pazocal{O}(\delta) $ terms in Eq.~\eqref{eq:appA-3} appear multiplied by $ \sqrt{\rho_T} $. Taking now the trace of Eq.~\eqref{eq:appA-3} one immediately sees that $ \tr y = \pazocal{O}(\delta^2) $ and hence, $ \mathbb{F}(\rho_T,\rho_{T+\delta}) = 1 + \pazocal{O}(\delta^2) $, as we wanted to verify.

\section{Low-temperature scaling of the QFI in the CLM}
\label{app:Ohmic_CL}

\subsection{Preliminaries}

In this appendix we will rigorously prove Eq.~\eqref{eq:scalings}. Recall that this refers to the low-temperature scaling of $ \pazocal{F}_T $ for a harmonic probe coupled to an equilibrium sample through an Ohmic spectral density with an arbitrary high-frequency cutoff function, as introduced in Eq.~\eqref{eq:SD_shape}. Essentially, we shall perform an asymptotic analysis on the definition of the QFI through Eqs.~\eqref{eq:qfi} and \eqref{eq:fidelity_covariance}, where we will insert the closed-form expressions for the covariances given in Eqs.~\eqref{eq:covariances}-\eqref{alph}. 

Let us start by writing down the Taylor expansions
\begin{subequations}\label{sp} 
	\bea \label{sp1}
	[\sigma_{T+\delta}]_{11} &=& [\sigma_T]_{11} + a_1 \delta T + b_1 \delta^2 + \pO(\delta^3),
	\\ \label{sp2}
	[\sigma_{T+\delta}]_{22} &=& [\sigma_T]_{22} + a_2 \delta + b_2 \delta^2 + \pO(\delta^3).~~~~
	\eea
\end{subequations}
The Uhlmann fidelity between $ \sigma_T $ and $ \sigma_{T+\delta} $ is thus
\bea \nonumber
\mathbb{F}(\sigma_T,\sigma_{T+\delta}) \! = \! 1 \! - \! \frac{a_1 a_2 + 2 [\sigma_T]_{11}^2 a_2^2 + 2 [\sigma_T]_{22}^2 a_1^2}{16 [\sigma_T]_{11}^2 [\sigma_T]_{22}^2 - 1} \delta^2 \! + \! \pO(\delta^3),
\eea
which leads to
\bea \label{fish1}
\pazocal{F}_T = 4 \frac{a_1 a_2 + 2 [\sigma_T]_{11}^2 a_2^2 + 2 [\sigma_T]_{22}^2 a_1^2}{16[\sigma_T]_{11}^2 [\sigma_T]_{22}^2 - 1}.
\eea
This is a very convenient expression, as it does not involve the second-order coefficients in Eqs.~\eqref{sp}. The problem of finding the low--$ T $ scaling $ \pazocal{F}_T $ is thus reduced to calculating the low-temperature expansions of $[\sigma_T]_{ii}$ (from where $ a_i(T) = d[\sigma_T]_{ii}/dT $). We will need to adopt two different strategies for the proof, for the cases $ \omega_0 \neq 0 $ [cf. Appendix \ref{app:scaling_trapped}] and $ \omega_0 = 0 $ [cf. Appendix \ref{app:scaling_free}], respectively.

\subsection{Probe with bare frequency $\omega_0 > 0$}\label{app:scaling_trapped}

Let us start with $[\sigma_T]_{11}$, which is given by \eqref{eq:q0}. Using the identity $ \coth{(x/2)}=1+2/(e^x-1) $, this covariance can be rewritten as
\bea \label{sig111}
[\sigma_T]_{11} = \sigma_1 + \frac{2}{\pi}\int_0^\infty d\omega\frac{J(\omega)}{|\alpha(\omega)|^2}\frac{1}{e^{\beta\omega}-1},
\eea
where we have defined
\bea \label{sigma1}
\sigma_1 \coloneqq \frac{1}{\pi}\int_0^\infty d\omega\frac{J(\omega)}{|\alpha(\omega)|^2}.
\eea

For convenience, we shall switch to the dimensionless parameters $ \tilde{\omega} \coloneqq \omega/\omega_c $ and $ \tilde{T}\coloneqq T/\omega_c $, normalized by the cutoff frequency $ \omega_c $. For a generic spectral density (as that of Fig.~\ref{fig:TIHC_spectral_dens}), $ \omega_c $ can be fixed from the maximization of $ J(\omega) $. After the transformation, Eq.~\eqref{sig111} turns into
\bea \label{sig112}
\sigma_{11}=\sigma_1 + \frac{2\omega_c}{\pi}\int_0^\infty d\tilde{\omega}\frac{J(\tilde{\omega}\omega_c)}{|\alpha(\tilde{\omega}\omega_c)|^2}\frac{1}{e^{\tilde{\omega}/\tilde{T}}-1}.
\eea
Hereafter, we shall drop the subscript $ T $ and the brackets in $ \sigma $ for simplicity of notation. We may split the integral in Eq.~\eqref{sig112} as $ \int_0^\infty \rightarrow \int_0^\xi + \int_{\xi}^{\infty} $, which leads us to
\begin{align}\label{eq:small_w}
\sigma_{11} = \sigma_1 + \frac{2\omega_c}{\pi}\int_0^\xi d\tilde{\omega}\frac{J(\tilde{\omega}\omega_c)}{|\alpha(\tilde{\omega}\omega_c)|^2}\frac{1}{e^{\tilde{\omega}/\tilde{T}}-1} + \sigma_1\pazocal{O}(e^{-\xi/\tilde{T}}),
\end{align}
where the last term encapsulates the fact that $ \int_\xi^\infty d\tilde{\omega}\frac{J(\tilde{\omega}\omega_c)}{|\alpha(\tilde{\omega}\omega_c)|^2}\frac{1}{e^{\tilde{\omega}/\tilde{T}}-1} < \frac{1}{e^{\xi/\tilde{T}}-1}\sigma_1 $. We can always choose $ \xi $ to be small but scale with temperature so that $ \pazocal{O}(e^{-\xi/\tilde{T}}) \rightarrow 0 $ exponentially when $ \tilde{T} \rightarrow 0 $ (e.g., $ \xi = \tilde{T}^{1/2} \ll 1 $), which entails that, in order to study the low--$ T $ scaling of $ \sigma_{11} $, it suffices to expand the integrand of Eq.~\eqref{eq:small_w} around $ \tilde{\omega} = 0 $. To do so, recall that $ \alpha(\omega) \coloneqq \omega_0^2+\omega_R^2-\omega^2 - \chi(\omega)-\mathrm{i} J(\omega) $ and that $\chi(\omega) \coloneqq \frac{1}{\pi} \text{P} \int_{-\infty}^\infty d\omega'\tilde{J}(\omega')/(\omega'-\omega)$. Evaluating the principal value in this latter definition yields
\begin{multline} 
\pi \chi (\tilde{\omega} \omega_c) = \int_0^\infty d\tilde{\omega}'\frac{J(\tilde{\omega}'\omega_c)}{\tilde{\omega}'+\tilde{\omega}}\\ 
+\lim_{\epsilon\to 0}\left[ \int_0^{\tilde{\omega}-\epsilon} d\tilde{\omega}'\frac{J(\tilde{\omega}'\omega_c)}{\tilde{\omega}'-\tilde{\omega}} + \int^{\infty}_{\tilde{\omega}+\epsilon} d\tilde{\omega}'\frac{J(\tilde{\omega}'\omega_c)}{\tilde{\omega}'-\tilde{\omega}} \right].
\label{rexi}\end{multline}
Note that, in the first integral, we have used the fact that $ J(\omega) $ extends to negative frequencies as an odd function [cf. remark below Eq.~\eqref{alph}]. Let us work with the dimensionless spectral density $ \tilde{J}(\tilde{\omega})\coloneqq J(\omega_c \tilde{\omega})/(\gamma\omega_c) $ which, in the notation of Eq.~\eqref{eq:SD_shape}, would amount to $ \tilde{J}(\tilde{\omega}) = \tilde{\omega} f(\tilde{\omega}) = \pazocal{O}(\tilde{\omega})\ll 1 $. This brings Eq.~\eqref{rexi} into the form
\begin{multline} \label{rex1}
\frac{\pi\chi(\tilde{\omega} \omega_c)}{\gamma\omega_c} = \int_{\tilde{\omega}}^\infty \frac{dx}{x} \tJ(x-\tilde{\omega}) 
\\ 
+\lim_{\epsilon\to 0}\left[ \int_\epsilon^\infty  \frac{dx}{x} \tJ(\omega+x) - \int^{\omega}_{\epsilon} \frac{dx}{x} \tJ(\omega-x) \right].
\end{multline}
We can further split the first integral \eqref{rex1} as $ \int_\epsilon^\omega + \int_\omega^\infty $ to get
\begin{multline}\label{rex2}
\frac{\pi\chi(\tilde{\omega} \omega_c)}{\gamma\omega_c} = \int_{\tilde{\omega}}^\infty \frac{dx}{x} \left[\tJ(x+\tilde{\omega})+\tJ(x-\tilde{\omega})\right] 
\\ 
+\lim_{\epsilon\to 0} \int_\epsilon^{\tilde{\omega}}  \frac{dx}{x} \left[\tJ(x+\tilde{\omega}) - \tJ(x-\omega) \right].
\end{multline}
Notice that, since $\tilde{\omega}\ll 1$, the second line of Eq.~\eqref{rex2} can be evaluated by expanding the spectral density around $ x = 0 $ as $\tJ(\tilde{\omega}+x)=\tJ(\tilde{\omega})+\tJ'(\tilde{\omega})x + \pO(x^2)$, where the primes denote derivatives. In particular, we have
\bea
\lim_{\epsilon\to 0} \int_\epsilon^{\tilde{\omega}}  \frac{dx}{x} \left[\tJ(x+\tilde{\omega}) - \tJ(x-\tilde{\omega}) \right] = 2\tilde{\omega} + \pazocal{O}(\tilde{\omega}^2),~~~
\label{rex3}\eea
since $ \tJ'(\tilde{\omega}) = \tJ'(0) + \pO(\tilde{\omega}) $ and $ \tJ'(0) = f(0) = 1 $. When it comes to the first term of Eq.~\eqref{rex2}, we may proceed similarly; Taylor-expanding the integrand around $ \tilde{\omega} = 0 $ this time, yields
\begin{multline}\label{blah}
\int_{\tilde{\omega}}^\infty \frac{dx}{x} \left[ \tJ(x+\tilde{\omega}) + \tJ(x-\tilde{\omega})\right] 
\\ = 2 \int_{\tilde{\omega}}^\infty \frac{dx}{x} \tJ(x) + \pO\left( \tilde{\omega}^2 \int_{\tilde{\omega}}^\infty \frac{dx}{x} \tJ''(x)\right).
\end{multline}
In turn, the first integral in Eq.~\eqref{blah} can be cast as
\begin{multline}\label{kalkun}
\int_{\tilde{\omega}}^\infty \frac{dx}{x} \tJ( x) = \int_0^\infty \frac{dx}{x} \tJ( x) - \int_0^{\tilde{\omega}} \frac{dx}{x} \tJ(x) 
\\ = \frac{\pi}{2} \frac{\omega_R^2}{\gamma\omega_c} - \tilde{\omega} + \pO(\tilde{\omega}^2 ),
\end{multline}
where we have used $ \tilde{J}(x) = \tilde{J}'(0) x + \pazocal{O}(x^2) $. To analyze the second term in Eq.~\eqref{blah}, we observe that
\begin{multline}\label{xutor}
\int_{\tilde{\omega}}^\infty \frac{dx}{x} \tJ''(x) = \int_{\tilde{\omega}}^1 \frac{dx}{x} \tJ''(x) + \int_1^\infty \frac{dx}{x} \tJ''(x)
\\ = 
\int_{\tilde{\omega}}^{1} \frac{dx}{x} \tJ''(x) + \pO(1).
\end{multline}
The fact that the second integral is $\pO(1)$ follows from $ \tJ''(x) = 2 f'(x) + x f''(x)$ and the requirement that $ f(x) $ should be a well-behaved function of $ x $ decaying rapidly for $ x > 1 $. Due to the \textit{shape} of an Ohmic spectral density, we may also write
\bea\label{xutor2}
\int_{\tilde{\omega}}^1 dx\frac{\tJ''(x)}{x} \leq -\pO(1)\ln{\tilde{\omega}}.
\eea
where $\leq$ signifies the fact that the left-hand side either scales as $\ln\tom$ or slower. Indeed, whenever $f(x)=f(0)+\pO(x)$ (e.g., when $f(x)=e^{-x}$), the left-hand side in Eq.~\eqref{xutor2} scales as $\ln\tom$, whereas if $f(x)=f(0)+\pO(x^2)$ [e.g., when $f(x)=1/(1+x^2)$], it becomes $\pO(1)$. 

Combining Eqs.~\eqref{rex2}--\eqref{xutor2} finally leads to
\bea
\chi(\tilde{\omega} \omega_c) = \omega_R^2 + \gamma\omega_c\pO(\tilde{\omega}^2\ln{\tilde{\omega}}),
\eea
which allows to cast $ \alpha(\tilde{\omega}\omega_c) $ in Eq.~\eqref{alph} as
\bea\nonumber
\alpha(\tilde{\omega} \omega_c) = \omega_0^2 - \tilde{\omega}^2\omega_c^2 + \gamma\omega_c\pO(\tilde{\omega}^2\ln{\tilde{\omega}}) - \mathrm{i} \gamma \omega_c \pO(\tilde{\omega}).\label{eq:alpha_not_squared} \\ 
\eea
Working under the physically relevant assumption that $ \gamma \leq \pO(\omega_0) $, we thus have
\bea \label{fldir1}
|\alpha(\tom\omega_c)|^2 = \omega_0^4 + \omega_c^4
\pO (\Lambda(\tom)),
\eea
where
\bea \label{fldir2}
\Lambda(\tom) = \tom^4 + \tom_0^2 \tom^2 + \tilde{\gamma}\tom_0^2 \tom^2 \ln \tom + \tilde{\gamma} \tom^4 \ln \tom~~~~~
\eea
and $\tom_0\coloneqq\omega_0/\omega_c$ and $\tilde{\gamma}\coloneqq\gamma/\omega_c$. Note that $\Lambda(\tom)\ll\tom_0^4$ whenever $\tom\ll\tom_0$. 

Substituting Eq.~\eqref{fldir1} into Eq.~\eqref{sig112}, up to exponentially small terms [cf. Eq.~\eqref{eq:small_w}], we get
\bea \label{fldir3}
\sigma_{11} = \sigma_1 + \frac{2\gamma\omega_c^2}{\pi\omega_0^4} \int\limits_0^{\sqrt{\tT}} \! \frac{d\tom}{e^{\tom/\tT}-1}\frac{\tJ(\tom)}{1 + \frac{\pO(\Lambda(\tom))}{\tom_0^4}}.
\eea
Since $\tom$ is small on the whole interval of integration, we can use $\tJ(\tom) = \tom [1+\pO(\tom)]$. Furthermore, changing the integration parameter in Eq.~\eqref{fldir3} to $A\coloneqq\tom/\tT$, we obtain
\bea \label{fldir4}
\sigma_{11} = \sigma_1 + \frac{2\gamma T^2}{\pi\omega_0^4} \int\limits_0^{1/\sqrt{\tT}} \! dA \frac{A}{e^A - 1} \frac{1+\pO(A\tT)}{1 + \frac{\pO(\Lambda(A\tT))}{\tom_0^4}}.~~~~~
\eea

Let us now study $\sigma_{11}$ in the $T\ll\omega_0$ ($\tT \ll \tom_0$) limit. The analysis is slightly different for $\tom_0 = \pO(1)$ and $\tom_0 \ll 1$. In the first case, $\tT \ll \tom_0$ implies $\sqrt{\tT}\ll\tom_0$, which means that $\pO(\Lambda(A\tT))\ll\tom_0^4$. Hence, given that $\int_0^\infty dA\frac{A}{e^A-1}=\frac{\pi^2}{6}$, Eq.~\eqref{fldir4} yields
\begin{subequations}
	\begin{align}\label{fldir5}
	\sigma_{11} &= \sigma_1 + \frac{\pi \gamma}{3\omega_0^4} T^2 [1+o(1)].\\
	a_1 &= \frac{2\pi\gamma}{3\omega_0^4} T[1+o(1)].
	\end{align}
\end{subequations}

In the second case (namely, when $\tom_0 \ll 1$) , $\sqrt{\tT}$ is not necessarily much smaller than $\tom_0$. Note however, that $ \omega_0 \ll \omega_c $ might be considered somewhat ``exotic'', as it would allow very large-frequency environmental modes to be \textit{effectively} coupled to the sample. $ \omega_c\rightarrow \infty $ would also entail a diverging renormalization frequency (e.g., $ \omega_R^2 = \gamma\omega_c $ for an Ohmic-Lorentzian spectral density). On the contrary, the example of Fig.~\ref{fig:TIHC_spectral_dens} shows that the condition $ \omega_0 \lessapprox \omega_c $ [i.e. $ \tilde{\omega}_0 = \pazocal{O}(1) $] appears naturally even in \textit{large} generic physical systems. Of course, whenever $\sqrt{\tT}\ll\tom_0$, we revert to Eq.~\eqref{fldir5}. Otherwise, we must note that since $\tT/\tom_0 \ll 1$, one also has that $\tT/\tom_0\ll\sqrt{\tT/\tom_0} \ll 1$. Hence, defining $A_m \coloneqq \sqrt{\omega_0/T}\gg 1$, we see that $A_m\tT\ll \tom_0$. It is thus convenient to split the integral Eq.~\eqref{fldir4} as $\int_0^{A_m} + \int_{A_m}^{\tT^{-1/2}}$. The first part evaluates to $\frac{\pi^2}{6}[1+o(1)]$, whereas the second is $\pO\big(e^{-\sqrt{\omega_0/T}}\big)$, thereby showing that Eq.~\eqref{fldir5} holds for \textit{any} $\omega_0$, provided that $T\ll\omega_0$. 

Conducting an identical analysis for the variance of the momentum, $\sigma_{22}$, yields
\begin{subequations}
	\begin{align}\label{fisk2p}
	\sigma_{22} &= \sigma_2 + \frac{2\pi^3\gamma}{15 \omega_0^4} T^4 [1+o(1)]~~\text{and}\\
	a_2 &= \frac{8\pi^3\gamma}{15 \omega_0^4} T^3 [1+o(1)],
	\end{align}
\end{subequations}
where
\bea 
\sigma_2\coloneqq\frac{1}{\pi}\int_0^\infty d\omega\frac{\omega^2 J(\omega)}{|\alpha(\omega)|^2}.
\eea

Finally, substituting everything back into Eq.~\eqref{fish1} gives us
\bea
\pazocal{F}(T) = \frac{32\pi^2}{144\sigma_1^2\sigma_2^2-9}\frac{\sigma_2^2\gamma^2}{\omega_0^8}T^2[1 + o(1)]\propto T^2.~~~~~
\eea

\subsection{Probe with bare frequency $\omega_0=0$}\label{app:scaling_free}

The analysis in the previous subsection does not entirely apply to this case since, whenever $\omega_0=0$ the integrals defining $\sigma_1$ and $\sigma_2$ diverge. In order to calculate the QFI, we thus need to regularize these divergences. We can do so by noticing that for any finite sample (no matter how large) there always exists a minimal frequency $ \omega_{\min} > 0 $ (i.e., an ``infrared cutoff''). This entails that the integrals in Eqs.~\eqref{eq:covariances} should start from $ \omega_{\min} $. The infrarred cutoff should be sent to $ 0 $ before taking any other limit (e.g., $ \omega_0/T \ll 1 $), which is equivalent to taking the thermodynamic limit. Hence, in order to calculate $\pazocal{F}(T)$, we should fix $T$ and evaluate $\lim_{\omega_{\min}\to 0}\pazocal{F}(T,\omega_{\min})$.

Note from Eq.~\eqref{eq:alpha_not_squared}, that the leading term in $ \vert \alpha(\tilde{\omega}\omega_c) \vert $ for $ \omega\ll 1 $ and $ \omega_0 = 0 $ is
\bea
|\alpha(\tom\omega_c)|^2 = \omega_c^2\gamma^2 \tom^2 \pO(1).
\eea
For convenience, let us write the $\pO(1)$ above as $W^{-1}+o(1)$, with $W$ being some $T$--independent dimensionless constant, potentially depending on $\gamma$ and $\omega_c$. This $ W $ will absorb all other constants in what follows. We shall introduce as well $ 0 < \epsilon \ll 1 $ (e.g. $\epsilon=\tT^2$), and make the following splitting:
\bea
\sigma_{11} = \frac{\gamma\omega_c^2}{\pi}\left[\int_{\tilde{\omega}_{\min}}^\epsilon + \int_\epsilon^\infty\right]d\tom\frac{\tJ(\tom)}{|\alpha(\tom\omega_c)|^2}\frac{e^{\tom/\tT}+1}{e^{\tom/\tT}-1}.~~
\eea
The second integral is non-singular and hence is $\pO(1)$ with respect to the limit $\tilde{\omega}_{\min}\to 0$. In turn, the first integral may be re-arranged as
\bea \nonumber
\frac{1}{\pi\gamma}\int_{\tilde{\omega}_{\min}}^\epsilon \frac{d\tom}{\tom} \frac{1+\pO(\tom)}{W^{-1}+o(1)}\frac{2+\pO(\tom/\tT)}{\tom/\tT[1+\pO(\tom/\tT)]}.
\eea
Noting that $\tom/\tT\leq\epsilon/\tT\leq \tT\ll 1$ and absorbing the numerical constants into $ W $, brings us to
\begin{subequations}
	\begin{align}
	\sigma_{11} &=\frac{T W}{\gamma \omega_c} \int_{\tilde{\omega}_{\min}}^\epsilon d\tom\frac{1+o(1)}{\tom^2}=\frac{T W}{\gamma \omega_{\min}} [1+o(1)]. \\
	a_1 &= \frac{W}{\gamma \omega_{\min}}  [1+o(1)].
	\end{align}
\end{subequations}

We now turn to $\sigma_{22}$, and define $g(\tom)\coloneqq\frac{|\alpha(\tom\omega_c)|^2}{\gamma^2\omega_c^2\tom^2}$, so that we can write it as (see Eq.~\eqref{eq:p0})
\begin{align} \label{bosor1}
\sigma_{22} = \frac{\omega_c^2}{\pi\gamma}\int_0^\infty d\tom\frac{\tom f(\tom)}{g(\tom)} + \frac{2\omega_c^2}{\pi\gamma}\int_0^\infty d\tom\frac{\tom f(\tom)}{g(\tom)}\frac{1}{e^{\tom/\tT}-1}.
\end{align}
Noticing that $g(0)>0$ and keeping in mind that $\tom f(\tom)$ is a rapidly decaying function of $\tom$ for $\tom>1$, we conclude that the first integral converges. As before, we denote it $\sigma_2$. Coming to the second integral in Eq.~\eqref{bosor1}, let us change the integration variable to $A=\tom/\tT$ and split the resulting integral as $\int_0^\infty=\int_0^{1/\sqrt{\tT}}+\int_{1/\sqrt{\tT}}^\infty$. It is straightforward to see that the second part is $\pO\big(e^{-1/\sqrt{\tT}}\big)$. Hence,
\begin{align}
\sigma_{22} = \sigma_2 + \frac{2T^2}{\pi\gamma}\int\limits_0^{1/\sqrt{\tT}}d A \frac{f(A \tT)}{g(A \tT)}\frac{A}{e^A - 1} + \pO\big(e^{-1/\sqrt{\tT}}\big).
\end{align}
Now, since the argument of the functions $f$ and $g$ does not exceed $\sqrt{\tT}$ on the interval of the integration, we can employ the Taylor expansion to observe that $\frac{f(A \tT)}{g(A \tT)}=\frac{f(0)}{g(0)} + o(1)$. Noticing furthermore that $\frac{A}{e^A-1}$ is finite at $A=0$, we conclude that
\bea
\sigma_{22}=\sigma_2+\frac{T^2}{\gamma}W'[1+o(1)],
\eea
where $W'=\frac{f(0)}{g(0)}\int_0^\infty d A \frac{A}{e^A - 1} = \frac{\pi^2}{6} \frac{1}{g(0)}$ is a dimensionless and temperature-independent constant [recall that $f(0)=1$; in fact, it it easy to see that $g(0)=f(0)^2=1$]. The coefficient $ a_2 $ is thus
\bea
a_2 = 2\frac{T}{\gamma}W'[1+o(1)].
\eea
It is important to note that here $o(1)$ is with respect to $\tT$.

The expression for the quantum Fisher information, \eqref{fish1}, thus leads to
\bea\label{psel}
\pazocal{F}_T=\frac12\lim_{\omega_{\min}\rightarrow 0}\big(T^{-2}[1+o(1)]\big) = \frac12 T^{-2}[1+o(1)],~~~~
\eea

Interestingly, this scales as the QFI of a free particle. Indeed, taking $\epsilon_n\to 0$ limit in Eq.~\eqref{cien}, we see that the heat capacity of a free particle $ C_{\text{free}}=1 $ and hence its QFI is $\pazocal{F}_{\text{free}}(T) = T^{-2}$. Eq.~\eqref{psel} thus means that, although the probe itself is not free when $ \omega_0=0 $ (recall that $\omega_R > 0$), it is coupled very efficiently to the zero-frequency mode.

For the error bar of the temperature measurements, $\delta T$, we thus have
\bea
\frac{\delta T}{T} \leq \frac{1}{\sqrt{M}},
\eea
where $M$ is the number of independent trials. In other words, if one makes measurements on a single node of a gapless TIHC, the error bar $\delta T$ will scale as $T$.

\section{Spectrum of 1D harmonic chain} \label{app:chain_gap}

In the limit of large $ N $, the gap of a TIHC is given by $ \Delta^2 = \Omega_N^2 = \Omega^2 - 2 \sum_{n=1}^\infty (-1)^{n-1} G_n + \Xi(s) $, which follows immediately from Eq.~\eqref{eigsC2}. In this appendix we will study the size-scaling of the error $ \Xi(s) $. Namely, we will show that
\begin{align}
\Xi(s) = \left\lbrace
\begin{array}{c}
\pazocal{O}(N^{-2})~~~~~~~ \qquad s > 2. \\
\pazocal{O}(N^{-2}\ln{N}) \qquad s = 2. \\
\pazocal{O}(N^{-s})~~~~~~~ \qquad s < 2.
\end{array}\right.
\label{blah1}
\end{align}

We shall start by manipulating Eq.~\eqref{eigsC2} so as to bring it into a convenient form:
\begin{align*}
\Delta^2 &= \Omega^2 + 2 \sum_{n=1}^N G_n \cos{\frac{2\pi n N}{2N +1}} \\
&= \Omega^2 - 2 \sum_{n=1}^N (-1)^{n-1} G_n \cos{\frac{\pi n}{2N +1}} \\
&= \Omega^2 - 2 \sum_{n=1}^N (-1)^{n-1} G_n + 4 \sum_{n=1}^N (-1)^{n-1} \sin^2{\frac{\pi n}{4N + 2}} \\
&= \Omega^2 - 2 \sum_{n=1}^\infty (-1)^{n-1} G_n + \Xi(s),
\end{align*}
where
\begin{equation}
\Xi(s) = 4 \sum_{n=1}^N (-1)^{n-1} \sin^2{\frac{\pi n}{4 N +2}} + 2 \sum_{n=N+1}^\infty (-1)^{n-1} G_n.
\label{blah2}
\end{equation}
Recall that $ G_n = G/n^s $ and hence, the explicit $ s $--dependence of the error. For the second sum, we have that
\begin{multline} \nonumber
\big\vert\sum_{n=N+1}^\infty (-1)^{n-1} G_n\big\vert \\= \big\vert G_{N+1} - \sum_{n=N+2}^\infty(G_n-G_{n+1})\big\vert \leq G_{N+1} = \pO(N^{-s}),
\end{multline}

We will now turn our attention to the first sum in Eq.~\eqref{blah2} and rewrite it using $\sum_{n=1}^N (-1)^{n-1}X_n = \sum_{n=1}^{N/2}(X_{2n-1} - X_{2n})$. For odd $N$, the upper limit would be $\lfloor N/2 \rfloor$ and the ``surplus'' term $ X_N = \pO(N^{-s})$ would be grouped with the second sum. We thus ignore this subtlety in what follows. Using the relation $\sin^2{x} - \sin^2{y} = \sin{(x-y)}\sin{(x+y)}$ yields
\begin{multline}
\Xi(s) = 4\sum_{n=1}^{N/2}(G_{2n-1}-G_{2n})\sin^2{\frac{\pi (2n-1)}{4N+2}} - \\ 
4 \sum_{n=1}^{N/2} G_{2n}\sin\frac{\pi}{4N+2}\sin{\frac{\pi(4n-1)}{4N+2}} + \pO(N^{-s}).
\label{eq:error2}
\end{multline}
The first term in the right-hand side of Eq.~\eqref{eq:error2} can be upper-bounded by
\begin{equation}
4\pi^2\sum_{n=1}^{N/2}\frac{(G_{2n-1}-G_{2n})\,n^2}{(2N+1)^2} = \pazocal{O}(N^{-2})\times\sum_{n=1}^{N/2}\pazocal{O}(n^{1-s}),
\label{eq:bounded_error1}
\end{equation}
were we have used that $ \sin{x}\leq x $ and $(G_{2n-1}-G_{2n}) = \pO(n^{-s-1})$. The size-scaling of the sum in right-hand side of Eq.~\eqref{eq:bounded_error1} depends on the value of $ s $ as
\begin{equation}
\sum_{n=1}^{N/2} \pazocal{O}(n^{1-s})=\left\lbrace\begin{array}{c}
\pazocal{O}(1)~~~~~~~~~ \qquad s > 2, \\
\pazocal{O}(\ln{N}) \,~~~~~~~~~~~~ s = 2, \\
\pazocal{O}(N^{2-s})~~~ \qquad s < 2,
\end{array}\right.
\label{blah4}
\end{equation}
This can be seen by noticing that $ \sum_{n=2}^N\int_n^{n+1} dx\, x^{-a} \leq \sum_{n=2}^N n^{-a} \leq \sum_{n=1}^{N-1}\int_n^{n+1} dx\, x^{-a} $. Looking at the size scaling of both bounds when $ a > 1$, $ a = 1 $, and $ a < 1 $, respectively, leads to Eq.~\eqref{blah4}. From which Eq.~\eqref{blah1} follows immediately.

Let us also mention that the maximal normal frequency corresponds to $a=0$ in Eq.~\eqref{eigsC2}, and is given by
\bea
\Omega_{\max}^2 = \Omega^2 + 2\sum_{n=1}^N G_n.
\eea
It thus scales as $N^{1-s}$ when $0<s<1$ and as $\ln N$ when $s=1$ whereas, for $G_n$ decaying faster than $n^{-1}$, $\Omega_{\max}$ is of the same order of magnitude as $\Omega$. 

Finally, our simulations indicate that, for any $s\geq 1$, the coupling constants in the corresponding CLM scale with $N$ as
\bea \label{blah3}
g_n \sim  n\,N^{-3/2},
\eea
including the cases in which $n/N=\pO(1)$.

\section{Consistent discretization of a continuous CLM} \label{app:correct_limit}

As advanced in Sec.~\ref{sec:star_to_TICH}, we may discretize the reservoir in a Caldeira-Leggett model by putting a cap on the environmental frequencies $ \omega < \omega_{\max} $ and sampling a large number $ N $ of discrete modes $ \omega_n = \omega_{\max}/N $ from the range $ (0,\omega_{\max}) $. In this appendix, we will elaborate on \textit{how large} does the maximum frequency and the number of environmental modes need to be in order to ensure that the discretized model represents the original system faithfully. In particular, we shall consider a CLM with the Ohmic-Lorentzian spectral density $ J(\omega) = 2\gamma\omega_c^2 \omega/(\omega^2 + \omega_c^2) $.

In particular, we will require the renormalization frequency to be well approximated in the discretized model. That is,
\bea
\omega_R^2 = \sum_{n=1}^N \frac{g_n^2}{\omega_n^2} \simeq \int_0^\infty d\omega\,\frac{J(\omega)}{\pi\omega} 
\eea

Let us start by rewriting Eq.~\eqref{eq:discrete_couplings_CL} as
\bea \label{gmu1}
\frac{1}{\gamma\omega_c}\frac{g_n^2}{\omega_n^2} = \frac{\omega_c N}{\omega_{\max}\,\pi\,n} \int\limits_{\frac{\omega_{\max}}{N}(n - 1/2 )}^{\frac{\omega_{\max}}{N}(n + 1/2 )} d\omega\,J(\omega), 
\eea
which evaluates to

\begin{align} \label{gmu2}
\frac{1}{\gamma\omega_c}\frac{g_n^2}{\omega_n^2} = \frac{\omega_c N}{\omega_{\max}\,\pi\,n}\,\ln{\frac{(\frac{\omega_{\max}}{\omega_c N})^2( n +\frac{1}{2} )^2+1}{(\frac{\omega_{\max}}{\omega_c N})^2( n -\frac{1}{2} )^2+1}}. 
\end{align}
Let us introduce the parameter $ a \coloneqq \frac{\omega_{\max}}{N\omega_c} $, which becomes small in the continuous limit, and rewrite Eq.~\eqref{gmu2} as.
\begin{align}\label{ccumb1}
\frac{1}{\gamma\omega_c}\frac{g_n^2}{\omega_n^2} = \frac{1}{\pi a n}\ln{\left[1+\frac{2a^2 n}{1+a^2 n^2}\left(1-\frac{a^2(n-\frac14)}{1+a^2 n^2}\right)^{-1}\right]}.
\end{align}
We shall also define $ x \coloneqq 2a^2 n/(1+a^2 n^2) $ and $ y\coloneqq a^2(n-\frac14)/(1+a^2 n^2) $. Noticing that $n/(1+a^2 n^2)\leq (2a)^{-1}$, we conclude that both $ x $ and $ y $ are $ \leq \pazocal{O}(a) \ll 1 $, and that, e.g., $ \pazocal{O}(x^2) = \pazocal{O}(y^2) = x\pazocal{O}(y) = y\pazocal{O}(x) $. In particular, $ x (1-y)^{-1} = x + x\pazocal{O}(y) = x + \pazocal{O}(x^2) \ll 1$, which justifies the use of the Taylor expansions $ \ln(1+z) = z - (1/2)\,z^2 + \pazocal{O}(z^3) $ and $ (1-z)^{-1} = 1 + z + \pazocal{O}(z^2) $ in Eq.~\eqref{ccumb1}, which then becomes
\begin{multline}\label{ccumb2}
\ln{[1+x(1-y)^{-1}]}
= x + xy - \frac12(x + xy)^2 + \pazocal{O}(x^3).
\end{multline}
This leads to
\bea \nonumber
\frac{1}{\gamma\omega_c}\frac{g_n^2}{\omega_n^2} = \! \frac{2a}{\pi}\frac{1}{1+a^2 n^2} -\frac{1}{2\pi}\frac{a^3}{(1+a^2 n^2)^2} \! + \! \pO \! \left( \!\! \frac{n^2 a^5}{(1+a^2 n^2)^3} \!\! \right), \! ~
\eea
which, by noticing that $\frac{n^2 a^5}{(1+a^2 n^2)^3}\leq\frac{a^3}{(1+a^2 n^2)^2}$, can be written as
\bea \label{ccumb3}
\frac{\omega_R^2}{\gamma\omega_c} = \frac{2a}{\pi}\sum_{n=1}^N \left[\frac{1}{1+a^2 n^2} + \pO \left( \frac{a^3}{(1+a^2 n^2)^2} \right)\right].~~~~
\eea

Back to Eq.~\eqref{ccumb3}, we may cast the first sum as
\begin{equation}\label{eq:split-sum}
\sum_{n=1}^N \frac{1}{1 + a^2 n^2} = \sum_{n=1}^\infty \frac{1}{1 + a^2 n^2} - \sum_{n=N+1}^\infty \frac{1}{1 + a^2 n^2},
\end{equation}
and exploit the identity $ \sum_{n=1}^\infty (1+a^2 n^2)^{-1} = \frac{\pi}{2a}\coth{\frac{\pi}{a}}-\frac12 $ for the first part. Note that, since $ a \ll 1 $, we have $ \coth{\frac{\pi}{a}} = 1 + \pazocal{O}(e^{-\pi/a}) $. In turn, the second term of Eq.~\eqref{eq:split-sum} can be approximated by means of the Euler-Maclaurin formula, as
\begin{align*}
\sum_{n=N+1}^\infty \frac{1}{a^2 n^2 + 1} = \frac{1}{a}\left(\frac{\pi}{2}-\arctan\frac{\omega_{\max}}{\omega_c} \right) + \pO \left( \frac{\omega_c^2}{\omega_{\max}^2} \right).
\end{align*}
We shall now take $\omega_{\max}\gg\omega_c$ and make use of the expansion $ \arctan{x} = \pi/2 - x^{-1} + \pO(x^{-2}) $ for $ x\gg 1$. This gives $ \sum_{n=N+1}^\infty (a^2 n^2 + 1)^{-1} = \pO (\omega_c^2 N/\omega_{\max}^2) $ and hence
\begin{align} \label{ccumb5}
\sum_{n=1}^N \frac{2a/\pi}{a^2 n^2 + 1} = 1 - \frac{a}{\pi} + \pO\left( \frac{\omega_c}{\omega_{\max}}\right) + \pO\left( a \,e^{-\frac{\pi}{a}} \right).
\end{align}

Only the second term in Eq.~\eqref{ccumb3} remains unevaluated. We may proceed as follows:
\bea\label{ccumb6}
\sum_{n=1}^N && \pO \left( \frac{a^3}{(1+a^2 n^2)^2} \right) \leq C \sum_{n=1}^N \frac{a^3}{(1+a^2 n^2)^2}
\nonumber \\ 
&& \leq C \int\limits_0^N \frac{ a^3 dx}{(1+a^2 x^2)^2} =  C a^2 \int\limits_0^{Na} \frac{ dx}{(1+x^2)^2} = a^2 \pO(1),~~~~~~~
\eea
where $C$ is some constant and $ \int_0^{Na} dx \frac{1}{(1+x^2)^2} = \pi/2 + \pazocal{O}(\frac{1}{Na}) $ (recall that $ Na = \omega_c/\omega_{\max} \ll 1 $). Substituting Eqs.~\eqref{ccumb5} and \eqref{ccumb6} into Eq.~\eqref{ccumb3}, we finally arrive at
\bea
\omega_R^2=\gamma\omega_c\left[ 1 - \frac{\omega_{\max}}{\pi N \omega_c} + \pO \! \left( \! \frac{\omega^2_{\max}}{N^2 \omega_c^2} \! \right) + \pO \! \left( \! \frac{\omega_c}{\omega_{\max}} \! \right) \! \right] \! , ~~~~~~~
\eea
which coincides with the continuous-limit value $\omega_R^2=\gamma\omega_c$ only when the conditions
\bea \label{regime}
\quad N\gg\frac{\omega_{\max}}{\omega_c}\gg 1
\eea
are satisfied. This shows that one must require not only that $\Delta\omega = \frac{\omega_{\max}}{N}\ll \omega_c$, but also that $\omega_{\max}$ be much larger than $\omega_c$. Note that if any of these relations is broken, $\omega_R^2$ can significantly differ from its actual value---it is thus essential to take this subtlety into account when discretizing a CLM with an Ohmic-Lorentzian spectral density.


%

\end{document}